\documentclass{aastex631} 
\usepackage{arydshln}
\usepackage{booktabs}
\usepackage{multirow}
\usepackage{subfigure}
\usepackage{multirow}
\usepackage{epstopdf}
\usepackage[normalem]{ulem}
\usepackage[figuresright]{rotating}
\usepackage{longtable}
\usepackage{makecell}
\usepackage{array}
\usepackage{soul}
\usepackage{url}
\usepackage{natbib}
\usepackage{longtable}
\usepackage{chngcntr}
\useunder{\uline}{\ul}{}

\newcommand{\velunit}{km~s$^{-1}$}

\shorttitle{}
\shortauthors{Shi et al.}
\graphicspath{{./}{figures/}}
\renewcommand{\baselinestretch}{1.6}
\begin{document}

\title{The Influence of  Magnetic Complexity of Active Regions on Solar Wind Properties During Solar Cycles 23 and 24}

\correspondingauthor{Hui Fu}
\email{fuhui@sdu.edu.cn}

\author{Xinzheng Shi}
\affiliation{Shandong Key Laboratory of Optical Astronomy and Solar-Terrestrial Environment, Institute of Space Sciences, Shandong University, Weihai, Shandong, 264209, China}

\author{Hui Fu}
\affiliation{Shandong Key Laboratory of Optical Astronomy and Solar-Terrestrial Environment, Institute of Space Sciences, Shandong University, Weihai, Shandong, 264209, China}

\author{Zhenghua Huang}
\affiliation{Shandong Key Laboratory of Optical Astronomy and Solar-Terrestrial Environment, Institute of Space Sciences, Shandong University, Weihai, Shandong, 264209, China}

\author{Limei Yan}
\affiliation{Key Laboratory of Earth and Planetary Physics, Institute of Geology and Geophysics, Chinese Academy of Sciences, Beijing, 100029, China}

\author{Qi Liu}
\affiliation{Shandong Key Laboratory of Optical Astronomy and Solar-Terrestrial Environment, Institute of Space Sciences, Shandong University, Weihai, Shandong, 264209, China}



\author{Lidong Xia}
\affiliation{Shandong Key Laboratory of Optical Astronomy and Solar-Terrestrial Environment, Institute of Space Sciences, Shandong University, Weihai, Shandong, 264209, China}




\begin{abstract}
Linking solar wind properties to the activities and characteristics of its source regions can enhance our understanding of its origin and generation mechanisms. Using the Mount Wilson magnetic classification (MWMC), we categorize all active regions (ARs) between 1999 and 2020 into three groups: $\alpha$, $\beta$, and complex ARs. Subsequently, we classify the near-Earth AR solar wind into the corresponding three types based on the magnetic type of ARs. Our results show that $\alpha$, $\beta$, and complex ARs account for 19.99\%, 66.67\%, and 13.34\% of all ARs, respectively, while their corresponding AR solar wind proportions are 16.96\%, 45.18\%, and 37.86\%.
The properties of solar wind from different types of ARs vary significantly. As the magnetic complexity of ARs increases, the corresponding AR solar wind exhibits higher magnetic field strength, charge states, helium abundance ($A_{He}$), and first ionization potential (FIP) bias. Our results demonstrate that complex ARs are more effective at generating solar wind. Additionally, the strong magnetic fields and frequent magnetic activities in complex ARs can heat the plasma to higher temperatures and effectively transport helium-rich materials from the lower atmosphere to the upper corona.
\end{abstract}


\section{Introduction}\label{sec:intro}
The solar wind is a supersonic plasma flow originating from the Sun. Its origin, heating, and acceleration mechanisms are important issues in solar and space physics \citep{2016SSRv..201...55A,2017SSRv..212.1345C,2020JGRA..12526005V,2021LRSP...18....3V}. Investigating the connection between the properties of the solar wind and their source regions can enhance our understanding of these processes. Large-scale coronal structures, including coronal holes (CHs), active regions (ARs), and the quiet Sun (QS), are considered important source regions of the solar wind (see \cite{2016SSRv..201...55A} and the references therein). The magnetic field strength and topology vary across these regions, resulting in distinct solar wind properties from different source regions \citep{2015SoPh..290.1399F,2017ApJ...836..169F,2024ApJ...972...54S}. The plasma density and temperature in CHs are significantly lower than those of surrounding regions, and the magnetic field in CHs is predominantly unipolar and primarily associated with open magnetic field lines \citep{1977RvGSP..15..257Z,2002SSRv..101..229C}. CHs are the primary source regions of fast solar wind \citep{1973SoPh...29..505K,1977RvGSP..15..257Z,2003A&A...399L...5X,2005Sci...308..519T}. The solar wind originating from CHs exhibits faster proton speeds and a lower charge state ratio \citep{2015SoPh..290.1399F,2017ApJ...836..169F,2017ApJ...846..135Z,2024ApJ...972...54S}. Additionally, the solar wind originating from CHs has a higher helium abundance \citep{2004AdSpR..33..689O,2024ApJ...972...54S} and exhibits higher Alfv\'{e}nicity \citep{2015ApJ...805...84D,2020A&A...633A.166P}.
The source of slow solar wind is more complex, potentially originating from CH boundaries, pseudo-streamers, QS, and ARs and their boundaries, as suggested by backtracking and imaging data.

ARs are also an important source of the solar wind \citep{2002JGRA..107.1488N,2015SoPh..290.1399F,2016SSRv..201...55A,2021SoPh..296..116S,2024ApJ...972...54S}. Global magnetic field structure models predict the location of open magnetic field lines and suggest that the AR should be the source of the solar wind. Subsequent case studies have shown that ARs contain a large number of open magnetic field lines, which can serve as sources of the solar wind \citep{2002JGRA..107.1488N,2003ApJ...587..818W,2005A&A...432L...1W}. \cite{1999JGR...10416993K} used the interplanetary scintillation tomography method and found that the solar wind originates from unipolar magnetic field regions on the side of ARs. Observation data from the Hinode satellite shows a continuous plasma outflow at the edges of ARs. The outflow has Doppler velocities ranging from 20 to 50 \velunit. This persistent outflow material contributes at least in part to the slow solar wind \citep{2007Sci...318.1585S,2008ApJ...676L.147H,2008ApJ...686.1362D,2015AGUFMSH54B..06B,2021SoPh..296...47T}. The remote measured first ionization potential (FIP) bias in the AR upflow regions matches the FIP bias of the solar wind measured at 1 AU a few days later, further confirming the contribution of the ARs to the solar wind \citep{2011ApJ...727L..13B,2020A&A...640A..28S}. Statistical studies also indicate that the contribution of ARs to solar wind should not be overlooked. During the solar maximum, about half of the ecliptic solar wind originates from ARs \citep{2015SoPh..290.1399F,2024ApJ...972...54S}. \cite{2021SoPh..296..116S} used the potential field source surface (PFSS) model, combined with solar photospheric magnetic field maps from the past four solar cycles, to track the footpoint distribution of open magnetic field lines at the photosphere. This method was employed to evaluate the contribution of ARs to the total solar wind. The statistical results show that during solar maximum, the contribution of ARs to all solar wind ranges from 30\% to 80\%. Compared to solar wind originating from CHs, solar wind from ARs is generally associated with lower speeds, lower helium abundance, and higher charge state ratios. Furthermore, solar wind from ARs shows higher FIP bias than the solar wind from CHs \citep{2004SoPh..223..209L,2015SoPh..290.1399F,2017ApJ...846..135Z,2024ApJ...972...54S}.

The magnetic field structures of sunspot groups in ARs exhibit significant differences. An active region is a three-dimensional area in the solar atmosphere controlled by magnetic fields. Sunspots, which are the core of ARs, are the most prominent features in the solar photosphere. Many studies use the magnetic complexity of ARs as a measure of their activity \citep{1901ApJ....13..260C,1938ZA.....16..276W,1990SoPh..125..251M,2016JSWSC...6A...3M}. The Mount Wilson (or Hale) magnetic classification (MWMC) for sunspot groups, proposed by Hale, has been in use for over a century \citep{1919ApJ....49..153H}. The MWMC is a method of classifying sunspot magnetic characteristics based on their magnetic structure. This classification scheme provides a simple way to describe the magnetic complexity of ARs. The scheme primarily includes three complexity classes: unipolar ($\alpha$), bipolar ($\beta$), and multipolar ($\gamma$). Since 1919, the original classification scheme has undergone several revisions. The most important revision occurred in 1960 when \cite{1960AN....285..271K} introduced a new complex type, $\delta$. Table \ref{tab1} shows the classification rules for grouping the magnetic complexity of ARs according to MWMC.

\begin{table}[htbp]
  \tiny
  \centering
  \caption{The Mount Wilson magnetic classification rules for categorizing ARs based on their magnetic complexity. (Referenced the p.80 in \cite{2015LRSP...12....1V} and the Table 1 in \cite{2019A&A...629A..45N})}
    \begin{tabular}{cl}
    \toprule
    \toprule
    \textbf{Class symbol} & \textbf{Classification rules} \\
    \midrule
    \textbf{$\alpha$}     & A unipolar sunspot group \\
    \midrule
    \textbf{$\beta$}     & A bipolar sunspot group with a distinct boundary between polarities \\
    \midrule
    \textbf{$\gamma$}     & A complex active region with mixed positive and negative polarities that it cannot be classified as a bipolar group \\
    \midrule
    \textbf{$\beta\gamma$}    & A bipolar sunspot group with no continuous boundary between opposite polarities \\
    \midrule
    \textbf{$\delta$}     & A complex sunspot group with opposite polarity umbrae within the same penumbra \\
    \midrule
    \textbf{$\beta\delta$}    & A $\beta$-type sunspot group containing one or more $\delta$ spots \\
    \midrule
    \textbf{$\beta\gamma\delta$}   & A $\beta\gamma$-type sunspot group containing one or more $\delta$ spots \\
    \midrule
    \textbf{$\gamma\delta$}    & A $\gamma$-type sunspot group containing one or more $\delta$ spots \\
    \bottomrule
    \end{tabular}%
  \label{tab1}%
\end{table}%

The MWMC has a long history, with an intuitive and relatively simple categorization, and it covers nearly all observed ARs. Therefore, this classification scheme is very helpful for the study of ARs and their associated activities. The complexity of sunspot groups, as classified by the MWMC, increases in the following order: $\alpha$, $\beta$, $\gamma$, $\beta\gamma$, $\delta$, $\beta\delta$, $\beta\gamma\delta$, and $\gamma\delta$ \citep{2017ApJ...834..150Y}. \cite{1987KFNT....3....7C} introduced a structural parameter, the effective distance ($d_{E}$), which can be used to quantify the magnetic complexity of ARs. This parameter can quantitatively describe the degree of isolation or mutual penetration between the two polarities in an AR. \cite{2007A&A...462.1121G} calculated the average $d_{E}$ values for different types of ARs classified according to MWMC. They found that the average $d_{E}$ for $\beta$ ARs was 0.65 $\pm$ 0.228, for $\gamma$ ARs was 0.81 $\pm$ 0.326, and for $\delta$ ARs was 1.34 $\pm$ 0.789. Their results show that as the magnetic complexity of ARs increases, the average $d_{E}$ also increases. Specifically, the $d_{E}$ value increases in accordance with the order of ARs classified by the MWMC \citep{2007A&A...462.1121G,2007AdSpR..39.1773G,2010MNRAS.405..111G}.

The numbers and proportions of ARs with different magnetic types vary with the solar cycle. \cite{2016ApJ...820L..11J} analyzes the evolution of various magnetic types of ARs from 1992 to 2015. The study identifies a total of 5468 ARs during this period, with the magnetic type of each AR determined by the type at the time when its area reached the maximum. Among them, $\alpha$ ARs account for 19.46\%, $\beta$ ARs for 64.23\%, and complex types such as $\beta\gamma$, $\beta\delta$ and $\beta\gamma\delta$ account for 16.16\%. Other classifications account for less than 1\% of the total. The proportions of different types of ARs vary with solar activity. The proportion of $\alpha$ ARs generally remains around 20\% during solar maximum and decline phases. During solar minimum, this proportion increases to 30-40\%. The proportion of $\beta$ ARs typically constitutes 60-70\%. During solar maximum, the proportion can reach up to 80\%, and during solar minimum, it drops to about 60\%. Moreover, there is no significant difference in the latitudinal distribution of simple and complex ARs as the solar cycle evolves \citep{2016ApJ...820L..11J}. In addition, \cite{2019A&A...629A..45N} conducted a statistical analysis of the daily number of simple ARs ($\alpha$ and $\beta$) and complex ARs ($\beta\gamma$ and $\beta\gamma\delta$) between 1996 and 2018. They examined the evolution of the simple and complex ARs over the solar cycle. Their results showed that simple ARs are closely correlated with sunspot numbers (r = 0.99), while the peak of complex ARs is delayed by two years. Most complex ARs appear during the second peak of the solar cycle or later. As the solar cycle progresses into the latter half of the solar maximum and the early stages of the declining phases, the role of complex ARs becomes relatively more important.

There are significant differences in the solar flare occurrence rates and properties among different types of ARs classified according to the MWMC. Previous studies have extensively examined the relationship between flare productivity and AR characteristics \citep{1939ApJ....89..555G,1963MNRAS.126..123G,1968IAUS...35...33S,2000ApJ...540..583S,2011JGRA..11612108C,2012SoPh..281..639L,2014MNRAS.441.2208G,2017ApJ...834..150Y}. It is generally believed that intense flares are more likely to occur in ARs with large sunspot groups, strong magnetic fields, and complex magnetic configurations. \cite{1968IAUS...35...33S} classified sunspot regions from Wilson's magnetic maps using a scheme similar to Hale's. Their results showed that the higher the magnetic complexity of an AR, the higher the flare productivity. \cite{1994SoPh..149..105S} found that during SC 22, 96\% of X-class flares occur in $\delta$ sunspot groups, and the flare production ability of $\delta$ sunspot regions was closely related to their lifetime. \cite{2000ApJ...540..583S} pointed out that, in general, the larger the sunspot group area, the stronger the corresponding flare tends to be. However, for ARs with similar areas, those with $\beta\gamma\delta$ sunspot groups tend to produce stronger flares than other types of ARs. \cite{2014MNRAS.441.2208G} compiled data on X-ray solar flares and sunspot groups for 28 years from 1983 to 2011, and analyzed the dependence of different flare classes on the magnetic classification of their source regions. The results indicate that the stronger the flare, the higher the probability that it occurs in $\beta\gamma\delta$ sunspot groups. For instance, 83.34\% of X-class flares, 62.35\% of M-class flares, 43.18\% of C-class flares, and 25.47\% of B-class flares occurred in $\beta\gamma\delta$ sunspot groups. In addition to differences in flare production efficiency, the properties of flares also vary across different magnetic types of ARs. \cite{2017ApJ...834..150Y} found that flares from complex regions ($\beta\gamma\delta$ or $\gamma\delta$) have shorter durations and higher intensities compared to those from simpler regions ($\alpha$, $\beta$, or $\gamma$).

Coronal mass ejections (CMEs) are among the most violent eruptive phenomena on the Sun, and their properties and occurrence rates are also closely related to the types of ARs. \cite{2008ApJ...680.1516W} found that larger, stronger, and more complex ARs are more likely to produce high-speed CMEs. \cite{2007AdSpR..39.1773G} analyzed 86 flare-CME events in 55 ARs and found that as the magnetic complexity of the ARs increases, the speed of the CMEs also increases. Among 31 fast CMEs, 18 originated from $\beta\gamma\delta$ ARs, 8 from $\beta\gamma$ ARs, 2 from $\beta\delta$ ARs, and only 3 from simple $\beta$ ARs. \cite{2011JGRA..11612108C} concluded that the size, strength, and complexity of ARs significantly affect CME production efficiency. ARs that produced multiple CMEs were generally associated with sunspot types of at least $\beta\gamma$, while 90\% of the ARs without any CMEs had sunspot types of only $\alpha$ or $\beta$.

There are significant differences in the photospheric magnetic fields, large-scale magnetic configurations, and corresponding activities of ARs with different sunspots types. Whether there are differences in the solar wind properties corresponding to different types of  ARs are worth studying. The present study aims to classify the solar wind from ARs between 1999 and 2020 based on the magnetic types of their source regions. We will then statistically investigate the properties of the different types of AR solar wind during SCs 23 and 24. This work will deepen the understanding of how the characteristics and activities of source regions influence the properties of solar wind, further improving our understanding of key issues such as the origin, heating, and acceleration mechanisms of the solar wind. The data and analysis methods will be presented in Section 2. The Section 3 compares and discusses the differences for the solar wind originating from ARs with different magnetic types during SCs 23 (1999-2009) and 24 (2010-2020). Finally, the results and conclusions are summarized in Section 4.

\section{Data and analysis Methods} \label{sec:style}
The present study aims to investigate the properties of the solar wind originating from ARs associated with different sunspot types during SCs 23 and 24. The first step is to correlate the in-situ solar wind with the ARs. Here, the dataset established by \cite{2024ApJ...972...54S} is adopted. In the dataset the near-Earth solar wind is classified into AR, CH, and QS categories according to the source region types. \cite{2024ApJ...972...54S} improved the standard two-step mapping procedure by considering the initial acceleration process of the solar wind. This improved method can link the near-Earth solar wind to its source region more accurately. In the study, the near-Earth solar wind is traced back to the Sun with a time resolution of 12 hours, meaning that each 12-hour interval of in-situ solar wind data is treated as one solar wind parcel, which is then mapped to a single footpoint on the solar surface. The footpoints of the near-Earth solar wind are overplotted on the photospheric magnetic field and EUV synoptic images. Based on the characteristics of the source region, the solar wind is classified into three types: CH, AR, and QS wind. AR wind is defined as solar wind with footpoints located in the magnetically concentrated areas with National Oceanic and Atmospheric Administration (NOAA) AR numbers. For more details on boundary determination and classification of AR regions, please refer to \cite{2015SoPh..290.1399F,2017ApJ...836..169F,2018MNRAS.478.1884F} and \cite{2024ApJ...972...54S}.

The second step is to classify the ARs based on the magnetic types of sunspots. The daily information on ARs, including the region number, location, sunspot number, and sunspot type, is obtained from the Solar Region Summary\footnote{\url{https://www.swpc.noaa.gov/products/solar-region-summary}} (SRS). The SRS, compiled by the Space Weather Prediction Center (SWPC), is a joint product of NOAA and the United States Air Force (USAF), providing a detailed daily description of ARs observed on the solar disk. The SWPC compiles the SRS after analyzing reports from the USAF Solar Optical Observing Network (SOON). 
Based on these datasets, we use the MWMC to categorize ARs associated with near-Earth solar wind in \cite{2024ApJ...972...54S}. For simplicity, ARs are divided into three groups: $\alpha$, $\beta$, and complex types. The complex type includes $\beta\gamma$, $\beta\delta$, and $\beta\gamma\delta$ sunspots. Both $\alpha$ and $\beta$ sunspots exhibit simple and well-defined magnetic structures. Specifically, $\alpha$ sunspots are unipolar, while $\beta$ sunspots are bipolar with a clear separation between the two polarities. In contrast, the magnetic structures of other sunspot types are more complex. Additionally, certain sunspot types with complex magnetic structures (such as $\gamma$, $\delta$, $\beta\gamma$, $\beta\delta$, and $\beta\gamma\delta$) are relatively rare, making their individual statistical analysis less meaningful. Therefore, to ensure statistical significance, the commonly observed $\beta\gamma$, $\beta\delta$, and $\beta\gamma\delta$ sunspot types are grouped together and classified as complex types. The magnetic type of sunspots evolves over time. Therefore, for the classification of AR solar wind, the magnetic type we used corresponds to the day when the solar wind is traced back to the Sun. We applied the above approach to classify the ARs associated with AR solar wind from 1999 to 2020. Then the properties of solar wind originating from ARs with different magnetic types are investigated.

In addition, the parameters of the solar wind are mainly measured by the Wind\footnote{\url{https://cdaweb.gsfc.nasa.gov/index.html}} \citep{1995SSRv...71....5A} and Advanced Composition Explorer \citep[ACE\footnote{\url{http://www.srl.caltech.edu/ACE}},][]{1998SSRv...86....1S} spacecraft. The solar wind speed and helium abundance ($A_{He}$) are measured by the Solar Wind Experiment \citep[SWE,][]{1995SSRv...71...55O,2006JGRA..111.3105K} on the Wind. We use the 92-second resolution dataset and process it into hourly averages, converting the time resolution to one hour \citep{Ogilvie_Fitzenreiter_Lazarus_Kasper_Stevens_2021}. The interplanetary magnetic field is measured by the Magnetic Field Investigation \citep[MFI,][]{1995SSRv...71..207L,2013AIPC.1539..211K} on Wind. The charge state ($Q_{Fe}$, $O^{7+}/O^{6+}$) and the FIP bias (Fe/O) of the solar wind is provided by the Solar Wind Ion Composition Spectrometer \citep[SWICS,][]{1998SSRv...86..497G,2014ApJ...789...60S} on ACE. The ACE/SWICS changed its working mode on August 23, 2011, due to a hardware anomaly \citep{2016ApJ...826...10Z}. SWICS 1.1 is a reprocessed dataset covering the period from the early mission phase until the anomaly. It has a time resolution of one hour. It applies improved data analysis methods and more accurate calibrations to account for instrumental and statistical effects \citep{2014ApJ...789...60S}. SWICS 2.0 is generated after the anomaly. It uses a new analysis approach to correct for calibration and statistical issues under the new instrument conditions. Its native time resolution is two hours. We duplicate each 2-hour data value for the corresponding hourly values to match the 1-hour time resolution of the pre-anomaly data. That is, the 2-hour resolution measurement value is assigned to the corresponding two hours (the value is the same for the two hours). Due to the different calibration standards before and after the anomaly, SWICS 1.1 and SWICS 2.0 should not be confused\footnote{\url{https://izw1.caltech.edu/ACE/ASC/level2/ss2_l2desc.html}}. They have different measurement and statistical characteristics. Therefore, due to the ACE/SWICS anomaly, we divide the statistical analysis of solar wind properties into two separate periods: before and after the anomaly. This ensures that ACE/SWICS data before and after the anomaly are treated independently. For SC 23, we use the period from 1999 to 2009 (before the anomaly), and for SC 24, we use data from 2012 to 2020 (after the anomaly). These two periods cover the main phases of the respective solar cycles. Although the data before and after the anomaly are not suitable for direct combination in long-term trend analyses, they can still be analyzed separately to study the influence of source region magnetic activity on AR solar wind properties. This approach is valid because each dataset uses internally consistent calibration methods. The annual sunspot number is supplied by the Solar Influence Data Center of the Royal Observatory of Belgium\footnote{\url{http://www.sidc.be/silso/home}}.

\begin{table}[!htbp]
\scriptsize
\centering
\caption{The numbers and proportions of different types of ARs and AR solar wind between 1999-2020.}
\label{tab2}
\begin{tabular}{ccccccc}
\midrule
\midrule
\multicolumn{1}{c}{} & \multicolumn{3}{c}{The numbers (proportions) of ARs} & \multicolumn{3}{c}{The hourly counts (proportions) of AR solar wind} \\
\cmidrule(r){2-4} \cmidrule(r){5-7}
Type & $\alpha$ & $\beta$ & complex & $\alpha$ & $\beta$ & complex \\
\cmidrule(r){2-4} \cmidrule(r){5-7}
\midrule
1999 & 74 (18.93\%) & 284 (72.63\%) & 33 (8.44\%) & 312 (13.83\%) & 1584 (70.21\%) & 360 (15.96\%) \\
2000 & 88 (18.45\%) & 346 (72.54\%) & 43 (9.01\%) & 360 (15.96\%) & 1584 (70.21\%) & 312 (13.83\%) \\
2001 & 100 (21.14\%) & 317 (67.02\%) & 56 (11.84\%) & 24 (1.23\%) & 216 (11.04\%) & 1716 (87.73\%) \\
2002 & 98 (20.72\%) & 311 (65.75\%) & 64 (13.53\%) & 24 (1.17\%) & 660 (32.16\%) & 1368 (66.67\%) \\
2003 & 58 (19.66\%) & 195 (66.10\%) & 42 (14.24\%) & 192 (10.19\%) & 864 (45.86\%) & 828 (43.95\%) \\
2004 & 41 (22.78\%) & 105 (58.33\%) & 34 (18.89\%) & 204 (14.66\%) & 684 (49.14\%) & 504 (36.21\%) \\
2005 & 23 (17.83\%) & 79 (61.24\%) & 27 (20.93\%) & 204 (15.74\%) & 672 (51.85\%) & 420 (32.41\%) \\
2006 & 21 (23.86\%) & 54 (61.36\%) & 13 (14.77\%) & 228 (25.33\%) & 552 (61.33\%) & 120 (13.33\%) \\
2007 & 12 (25.53\%) & 29 (61.70\%) & 6 (12.77\%) & 216 (36.73\%) & 276 (46.94\%) & 96 (16.33\%) \\
2008 & 8 (26.67\%) & 22 (73.33\%) & 0 (0.00\%) & 108 (23.68\%) & 348 (76.32\%) & 0 (0.00\%) \\
2009 & 6 (20.69\%) & 23 (79.31\%) & 0 (0.00\%) & 12 (12.50\%) & 84 (87.50\%) & 0 (0.00\%) \\
\midrule
2010 & 20 (19.80\%) & 73 (72.28\%) & 8 (7.92\%) & 252 (30.43\%) & 384 (46.38\%) & 192 (23.19\%) \\
2011 & 42 (16.94\%) & 172 (69.35\%) & 34 (13.71\%) & 180 (9.62\%) & 504 (26.92\%) & 1188 (63.46\%) \\
2012 & 49 (19.68\%) & 170 (68.27\%) & 30 (12.05\%) & 156 (10.48\%) & 432 (29.03\%) & 900 (60.48\%) \\
2013 & 61 (20.40\%) & 181 (60.54\%) & 57 (19.06\%) & 264 (16.67\%) & 444 (28.03\%) & 876 (55.30\%) \\
2014 & 53 (16.88\%) & 177 (56.37\%) & 84 (26.75\%) & 468 (33.62\%) & 408 (29.31\%) & 516 (37.07\%) \\
2015 & 43 (19.20\%) & 145 (64.73\%) & 36 (16.07\%) & 324 (19.15\%) & 876 (51.77\%) & 492 (29.08\%) \\
2016 & 21 (14.29\%) & 120 (81.63\%) & 6 (4.08\%) & 168 (16.28\%) & 852 (82.56\%) & 12 (1.16\%) \\
2017 & 24 (34.78\%) & 40 (57.97\%) & 5 (7.25\%) & 468 (51.32\%) & 288 (31.58\%) & 156 (17.11\%) \\
2018 & 11 (28.95\%) & 26 (68.42\%) & 1 (2.63\%) & 84 (36.84\%) & 144 (63.16\%) & 0 (0.00\%) \\
2019 & 8 (38.10\%) & 11 (52.38\%) & 2 (9.52\%) & 144 (60.00\%) & 96 (40.00\%) & 0 (0.00\%) \\
2020 & 11 (27.50\%) & 28 (70.00\%) & 1 (2.50\%) & 168 (35.00\%) & 192 (40.00\%) & 120 (25.00\%) \\
\midrule
SC 23 & 529 (20.25\%) & 1765 (67.57\%) & 318 (12.18\%) & 1884 (12.45\%) & 7524 (49.72\%) & 5724 (37.83\%) \\
SC 24 & 343 (19.60\%) & 1143 (65.31\%) & 264 (15.09\%) & 2676 (22.78\%) & 4620 (39.32\%) & 4452 (37.90\%) \\
ALL & 872 (19.99\%) & 2908 (66.67\%) & 582 (13.34\%) & 4560 (16.96\%) & 12144 (45.18\%) & 10176 (37.86\%) \\
\midrule
\end{tabular}
\end{table}


\section{RESULTS AND DISCUSSION}
\subsection{The proportions of the three types of ARs and AR solar wind during SCs 23 and 24}
\begin{figure}[!h]
\hspace{-2.2cm}
\includegraphics[width=1.2\textwidth]{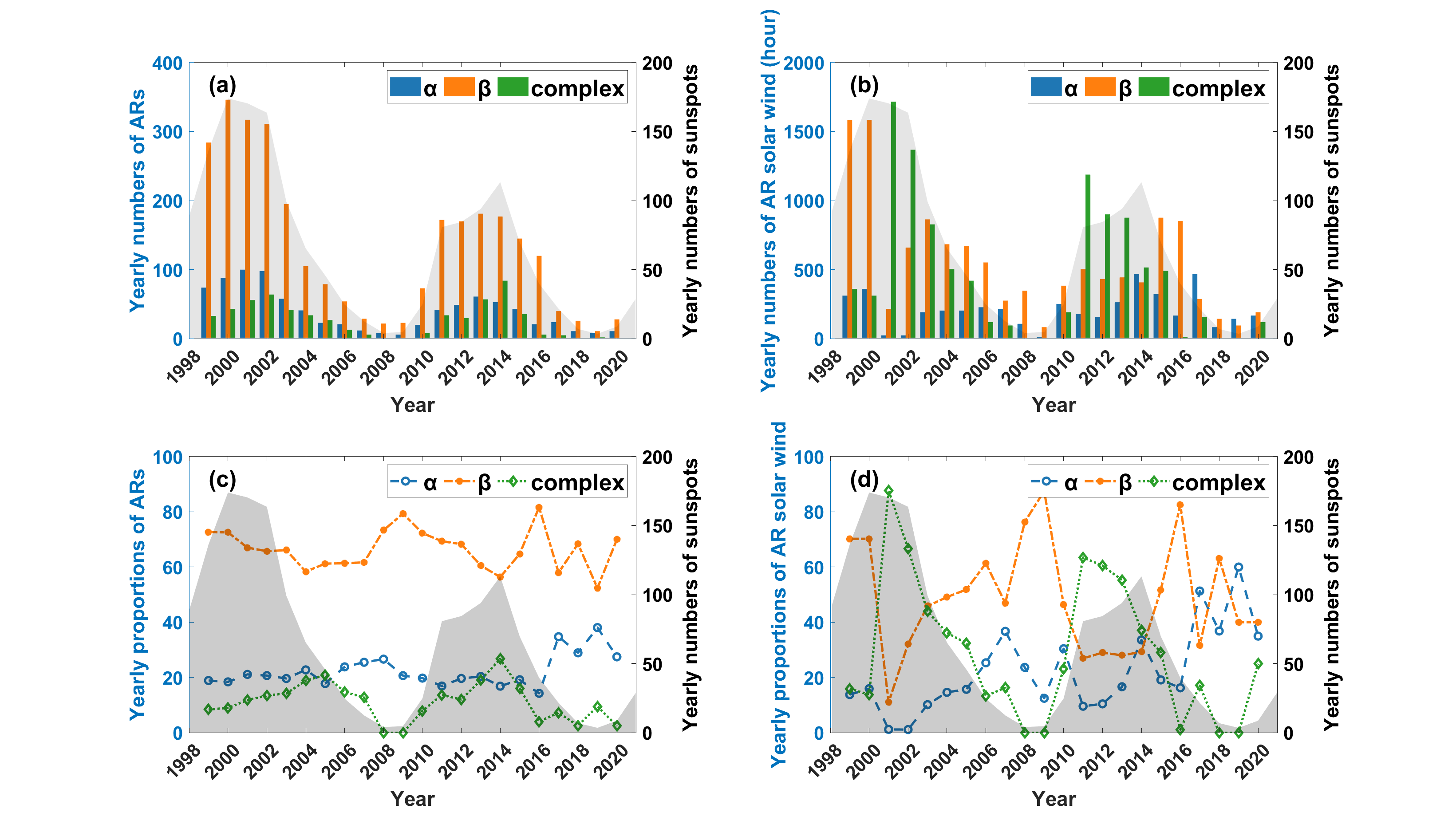}
\caption{The yearly numbers and proportions of different types of ARs and AR solar wind. The panel (a) represents the annual number of three types of ARs, and the panel (b) represents the yearly samples of the different types of AR solar wind. The bottom panel represents the annual proportions of each type of AR (panel (c)) and the annual proportions of three types of AR solar wind (panel (d)). The blue, orange and green represent $\alpha$, $\beta$, and complex ARs (panels (a) and (c)) and AR solar wind (panels (b) and (d)). The shaded area represents the annual sunspot numbers. The numbers and proportions of different types of ARs and AR solar wind show clear variations in response to solar activity.}
\label{year}
\end{figure}

Between 1999 and 2020, a total of 4,362 ARs classified as $\alpha$, $\beta$, and complex types are assigned unique numbers. For each AR, we choose its magnetic type on the day it reaches the maximum sunspot area. Table \ref{tab2} presents the numbers and proportions of the three types of ARs and AR solar wind during SCs 23 and 24. 

The yearly numbers and proportions of the three types of ARs and AR solar wind are presented in Figure \ref{year}. The numbers of different types of ARs exhibits significant fluctuations with the evolution of solar activity (see the panel (a) of Figure \ref{year}), this result is consistent with the statistical findings of \cite{2016ApJ...820L..11J} and \cite{2019A&A...629A..45N}. As solar activity weakens, the numbers of all three types of ARs decreases rapidly. Correspondingly, the annual proportions of the three types of ARs also vary with the changes in solar activity (see the panel (c) of Figure \ref{year}). The $\alpha$ ARs account for about 20\% each year and increase to about 30\% during solar minimum. The proportion of $\beta$ ARs remain between 60\% and 80\% each year. The proportion of complex ARs shows a significant change, reaching very low levels during solar minimum. The SC 23 has a much higher sunspot number than SC 24. During SC 23, the numbers of all three types of ARs are much higher than those during SC 24 (see panel (a) in Figure \ref{year}). The numbers of $\alpha$, $\beta$, and complex ARs during SC 23 (24) are 529 (343), 1765 (1143), and 318 (264), respectively. However, the proportions of the three types of ARs are similar between the two solar cycles. The proportions of $\alpha$, $\beta$, and complex ARs during SC 23 (24) are 20.25\% (19.60\%), 67.57\% (65.31\%), and 12.18\% (15.09\%), respectively. 

The hourly samples and proportions of the three types of AR solar wind also change significantly with sunspot numbers (see the panel (b) and (d) of Figure \ref{year}). The majority of AR solar wind originates from $\beta$ and complex ARs, while $\alpha$ ARs contribute less to the AR solar wind. Between 1999 to 2020, the proportions of $\alpha$, $\beta$, and complex AR solar wind are 16.96\%, 45.18\%, and 37.86\%, respectively. The hourly samples for the AR solar wind originating from $\alpha$, $\beta$, and complex ARs are 4560, 12144, and 10176, respectively. The contribution of different types of ARs to the solar wind is also significantly modulated by solar activity (see the panel (d) of Figure \ref{year}). As solar activity weakens, the proportion of complex AR solar wind decreases notably, while the proportion of $\alpha$ AR solar wind increases. During solar maximum, especially during the maximum phase of SC 23, the solar wind predominantly originates from $\beta$ and complex ARs. Their combined contribution is nearly 90\%.

It is worth noting that during SCs 23 and 24, which have significant differences in sunspot numbers, the contribution of the same type of ARs to the solar wind differs between the two solar cycles. The numbers of $\beta$ and complex AR solar wind during SC 23 are much higher than those during SC 24, while the number of $\alpha$ AR solar wind during SC 23 is lower than that during SC 24. The hourly samples of $\alpha$, $\beta$, and complex AR solar wind during SC 23 (24) are 1884 (2676), 7524 (4620), and 5724 (4452), respectively. During SC 23, the proportion of $\alpha$ ($\beta$) AR solar wind also significantly lower (higher) than during SC 24. But the proportion of complex AR solar wind remains similar across the both solar cycles. The proportions of $\alpha$, $\beta$, and complex ARs solar wind during SC 23 (24) are 12.45\% (22.78\%), 49.72\% (39.32\%), and 37.83\% (37.90\%), respectively.

Interestingly, the complex ARs, which account for only 13.34\% of all ARs, contribute 37.86\% to the AR solar wind. In contrast, although $\beta$ ARs make up the largest proportion (66.67\%), they contribute only 45.18\% to the AR solar wind. By dividing the hourly samples of AR solar wind by the number of ARs, it is found that the ratio for complex ARs is the highest, at 17.48, followed by $\alpha$ ARs at 5.23 and $\beta$ ARs at 4.18. This indicates that complex ARs making a greater relative contribution to the AR solar wind. This could be because the complex ARs undergo more frequent magnetic reconnection events, which enhance the release of plasma from closed magnetic loops into the AR solar wind. In contrast, the magnetic structure of $\beta$ ARs is relatively regular and resembles a dipole field with most magnetic field lines remaining closed, making it less likely to release plasma.

\begin{figure}[!h]
\hspace{-2.2cm}
\includegraphics[width=1.2\textwidth]{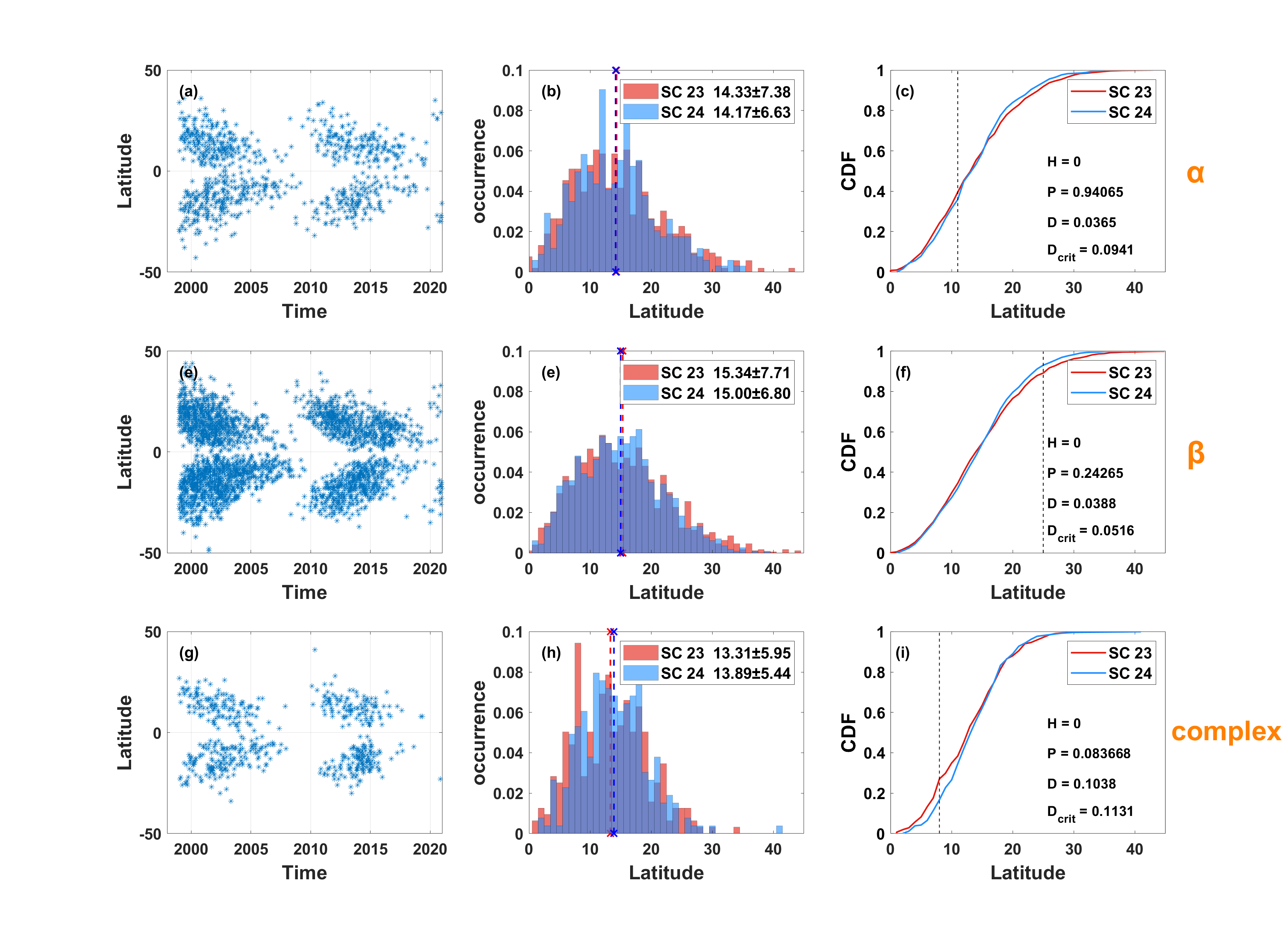}
\caption{The top, middle, and bottom rows respectively present the statistical results of the latitudinal distributions of all $\alpha$, $\beta$, and complex ARs from 1999 to 2020. The first column shows the temporal evolution of latitudes for the three types of ARs. The second column compares the latitudinal histograms of the three types of ARs during SC 23 (red) and SC 24 (blue). The third column presents the cumulative distribution function (CDF) curves of the latitudinal distributions for the same types of ARs between the two solar cycles, along with the results of the two sample Kolmogorov-Smirnov (K-S) test. The black vertical dotted line indicates the point where the maximum vertical deviation between the two curves, and the D is the difference between the two curves at that point. The results indicate that the latitudinal distributions of the three types of ARs are very similar, with no significant differences. Moreover, the latitudinal distributions of the same type of ARs are also similar between SC 23 and SC 24.}
\label{latitude}
\end{figure}
Figure \ref{latitude} shows the statistical results of the latitudinal distributions of the three types of ARs during the period from 1999 to 2020. The first column shows the temporal evolution of the latitudes of $\alpha$, $\beta$, and complex ARs. The second column displays histograms of their latitudinal distributions in SC 23 (red) and SC 24 (blue). The third column shows the cumulative distribution function (CDF) curves of the latitudinal distributions for the same type of ARs during SC 23 and SC 24, along with the results of the two sample Kolmogorov-Smirnov (K-S) test. The two sample K-S test can be used to determine whether two samples come from the same distribution. H, P, D, and critical D ($D_{\mathrm{crit}}$) are the key indicators of the two sample K-S test. The value of H indicates the result of the test: if H = 1, the null hypothesis is rejected at the given significance level (typically 0.05), suggesting that the two sample distributions are significantly different. If H = 0, the null hypothesis cannot be rejected. The P-value represents the significance probability of the test. It indicates the probability of observing the current sample difference (or a more extreme one) under the assumption that the null hypothesis is true. It is used to assess the evidence against the null hypothesis by comparing it with a predefined significance level (typically 0.05). A small P-value indicates strong evidence against the null hypothesis, suggesting that the two samples may come from different distributions. A large P-value indicates insufficient evidence to reject the null hypothesis, suggesting the distributions may not be significantly different. The D is defined by the maximum vertical deviation between the two curves. It quantifies the degree of distributional deviation. The larger the D value, the greater difference between the two distributions. The $D_{\mathrm{crit}}$ is the minimum value that the D must exceed in order for the test to be significant at the given significance level. Therefore, to reject the null hypothesis in a rigorous statistical sense, it is necessary that: D is greater than $D_{\mathrm{crit}}$ and P is less than the significance level. All results of the two sample K-S tests in this study are summarized in the Table \ref{tab4} in the appendix.

The statistical results indicate that the latitudinal evolution of the three types of ARs over time is highly consistent with the sunspot butterfly diagram (panels (a), (e), and (g)). According to the histograms, the latitudinal ranges of the different types of ARs are generally similar, with only the complex ARs showing a slightly narrower distribution. This suggests that there are no significant differences in the latitudinal distributions among different types of ARs. This result is consistent with the findings of \cite{2016ApJ...820L..11J}. Moreover, the latitudinal distributions of the same type of ARs show small differences between SC 23 and SC 24, with their histograms nearly overlapping. The average latitudes of $\alpha$, $\beta$, and complex ARs in SC 23 (SC 24) are 14.33$^{\circ}$ (14.17$^{\circ}$), 15.34$^{\circ}$ (15.00$^{\circ}$), and 13.31$^{\circ}$ (13.89$^{\circ}$), respectively. The results of the two sample K-S test on the latitudes of the same type of ARs in the two solar cycles (panels (c), (f), and (i) of Figure \ref{latitude}) show that all D values are smaller than $D_{\mathrm{crit}}$ and all P values are greater than 0.05; therefore, all H values are 0. This indicates that the CDFs of the latitudes of the same type of ARs between the two solar cycles do not differ significantly. Since the CDF describes the characteristics of the data distribution, this suggests that their latitudinal distributions are similar. Therefore, these results suggest that the latitudinal distribution of the same type of ARs is not significantly modulated by the amplitude of the solar cycle.


\subsection{The magnetic field strengths and speeds of the three types of AR solar wind during SCs 23 and 24}
\begin{figure}[!h]
\hspace{-2cm}
\setlength{\abovecaptionskip}{-0.5cm}
\includegraphics[width=1.2\textwidth]{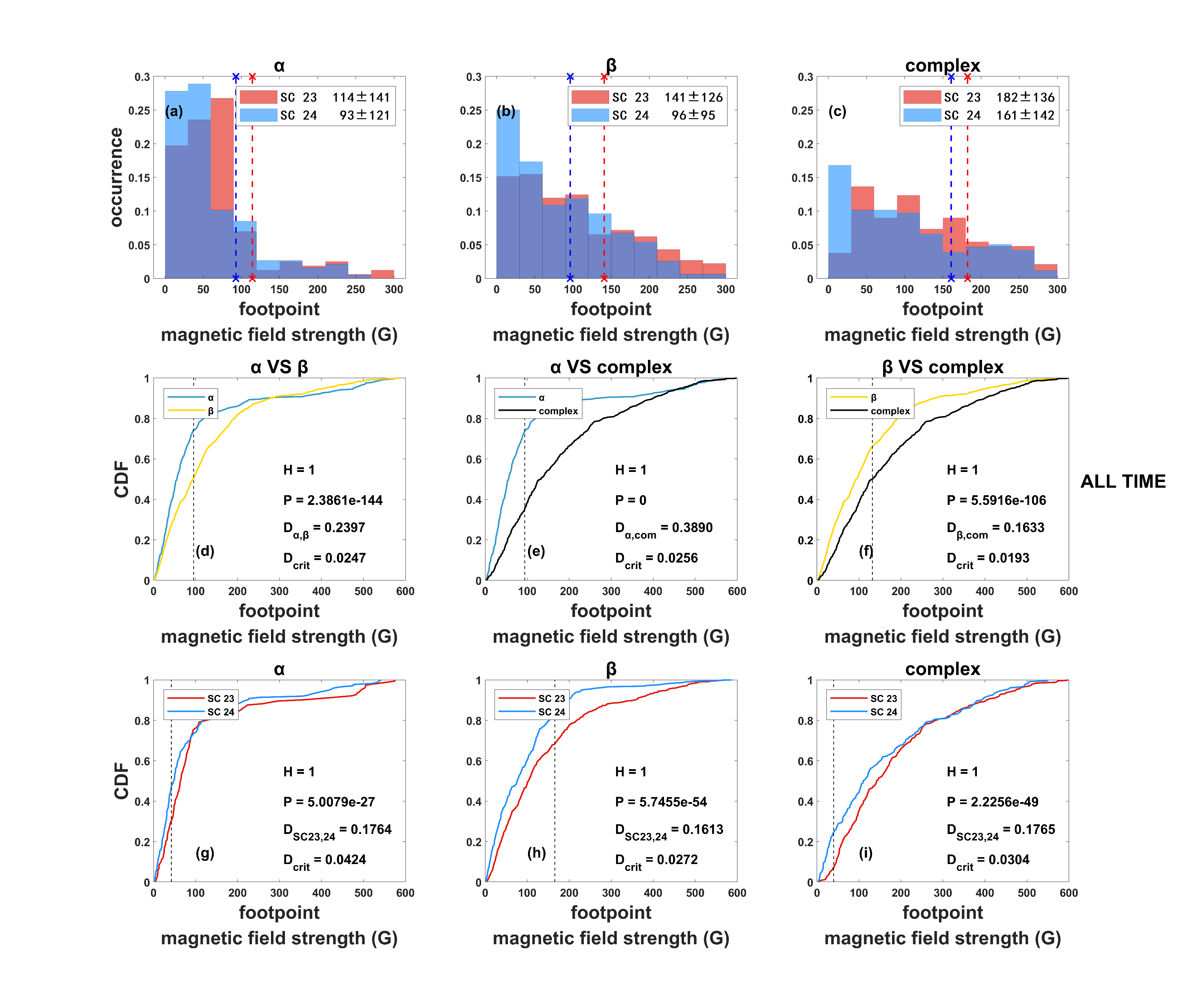}
\caption{The footpoint magnetic field strengths of $\alpha$, $\beta$, and complex AR solar wind during SCs 23 and 24. The first row shows the histograms of footpoint magnetic field strengths of the three types of AR solar wind during SC 23 (red) and SC 24 (blue). The second row presents the CDF curves of footpoint magnetic field strengths for different types of AR solar wind, combining data from SC 23 and SC 24, along with the two sample K-S test results between any two groups. The cyan, yellow, and black curves represent $\alpha$, $\beta$, and complex AR solar wind, respectively. The third row compares the CDF curves of footpoint magnetic field strengths for $\alpha$, $\beta$, and complex AR solar wind between SC 23 and SC 24. The footpoint magnetic field strength of complex AR solar wind is significantly higher than that of the other two types of AR solar wind.}
\label{footpoint}
\end{figure}
The footpoint magnetic field strengths of $\alpha$, $\beta$, and complex AR solar wind are presented in Figure \ref{footpoint}. The panels (a)-(c) of Figure \ref{footpoint} show the footpoint magnetic field strength histograms of the three types of AR solar wind during SCs 23 (red) and 24 (blue). The panels (d)-(f) of Figure \ref{footpoint} display the CDF curves of footpoint magnetic field strengths between any two types of AR solar wind. The panels (g)-(i) sequentially present the the CDF curves of footpoint magnetic field strengths for $\alpha$, $\beta$, and complex AR solar wind between SC 23 and SC 24. We also use the two sample K-S test to evaluate property differences between any two AR solar wind groups. The footpoint magnetic field strength of AR solar wind is influenced by both the magnetic complexity of the ARs and the amplitude of the solar cycle. The average footpoint magnetic field strengths for $\alpha$, $\beta$, and complex AR solar wind during SC 23 (24) are 114 (93), 141 (96), and 182 (161) Gauss, respectively. As the magnetic complexity of the source region increases, the footpoint magnetic field strength of AR solar wind also increases. The comparison of CDF curves also reveals significant differences in the footpoint magnetic field strengths among the three types of AR solar wind, the CDF curves of footpoint magnetic field strengths for complex AR solar wind are significantly lower than those of $\alpha$ and $\beta$ AR solar wind. This shows that the footpoint magnetic field strength of complex AR solar wind is markedly higher than that of $\alpha$ and $\beta$ AR solar wind. The corresponding two sample K-S statistic D ($D_{\mathrm{crit}}$) values are 0.2397 (0.0247), 0.3890 (0.0256), and 0.1633 (0.0193), respectively. All D values are significantly greater than their corresponding $D_{\mathrm{crit}}$ values, and all P values are well below 0.05; therefore, the H values of the two sample K-S test are all 1. Note that P = 0 (see the panel (e) of Figure \ref{footpoint}) occurs because the P-value is so small that it exceeds the numerical precision, not because the P-value is actually zero. This indicates that the P-value is extremely small, and it implies with very high confidence that the null hypothesis can be rejected. These results suggest that there are statistically significant differences in the distributions of footpoint magnetic field strengths among different types of AR solar wind. The CDF curves of footpoint magnetic field strengths for different types of AR solar wind during SC 23 generally lie below those during SC 24 (see panels (g)-(i) in Figure \ref{footpoint}). This is because during SC 23, which has higher sunspot numbers and stronger solar activity, the footpoint magnetic field strength of AR solar wind is significantly higher than that during SC 24 for the same types of AR solar wind.


\begin{figure}[!h]
\hspace{-2cm}
\setlength{\abovecaptionskip}{-0.5cm}
\includegraphics[width=1.2\textwidth]{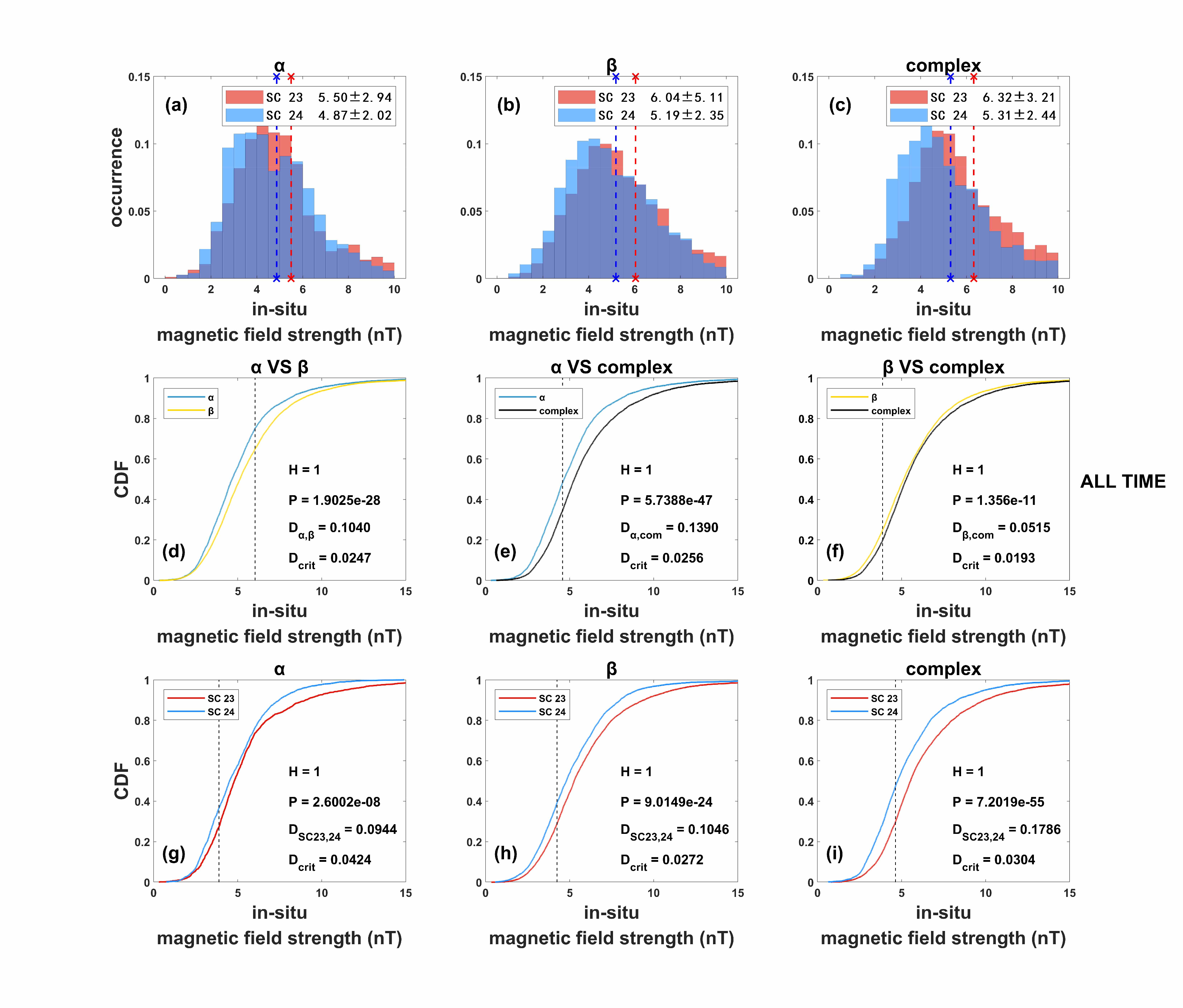}
\caption{The in-situ magnetic field strengths of $\alpha$, $\beta$, and complex AR solar wind. The layout is the same as Figure \ref{footpoint}. The higher the magnetic complexity of the AR, the higher the in-situ magnetic field strength of the solar wind from the AR.}
\label{B}
\end{figure}
The in-situ magnetic field strengths of $\alpha$, $\beta$, and complex AR solar wind are presented in Figure \ref{B}. The panels (a)-(c) of Figure \ref{B} present the statistical results of the in-situ magnetic field strengths of AR solar wind from three types of ARs during SCs 23 (red) and 24 (blue). The average in-situ magnetic field strengths for $\alpha$, $\beta$, and complex AR solar wind are 5.50 (4.87), 6.04 (5.19), and 6.32 (5.31) nT during SCs 23 (24), respectively. Similar to the footpoint magnetic field strengths of different types of AR solar wind, the results show that the in-situ magnetic field strengths of AR solar wind also increase with the magnetic complexity of the ARs. The comparison of CDF curves shows a relatively large difference between $\alpha$ AR solar wind and $\beta$ and complex AR solar wind. In contrast, the CDF curves for $\beta$ and complex AR solar wind are more similar (see the panels (d)-(f) of Figure \ref{B}). The corresponding D ($D_{\mathrm{crit}}$) values are 0.1040 (0.0247), 0.1390 (0.0256), and 0.0515 (0.0193), respectively. All D values are significantly greater than their corresponding $D_{\mathrm{crit}}$ values, and all P values are significantly smaller than 0.05. Therefore, all H values are 1. The two sample K-S test results demonstrate differences in the distributions of in-situ magnetic field strengths among different types of AR solar wind. For the same type AR solar wind, the in-situ magnetic field strengths during SC 23 are significantly higher than those during SC 24. This difference becomes more pronounced with increasing magnetic complexity of ARs.

\begin{figure}[!h]
\hspace{-2cm}
\setlength{\abovecaptionskip}{-0.5cm}
\includegraphics[width=1.2\textwidth]{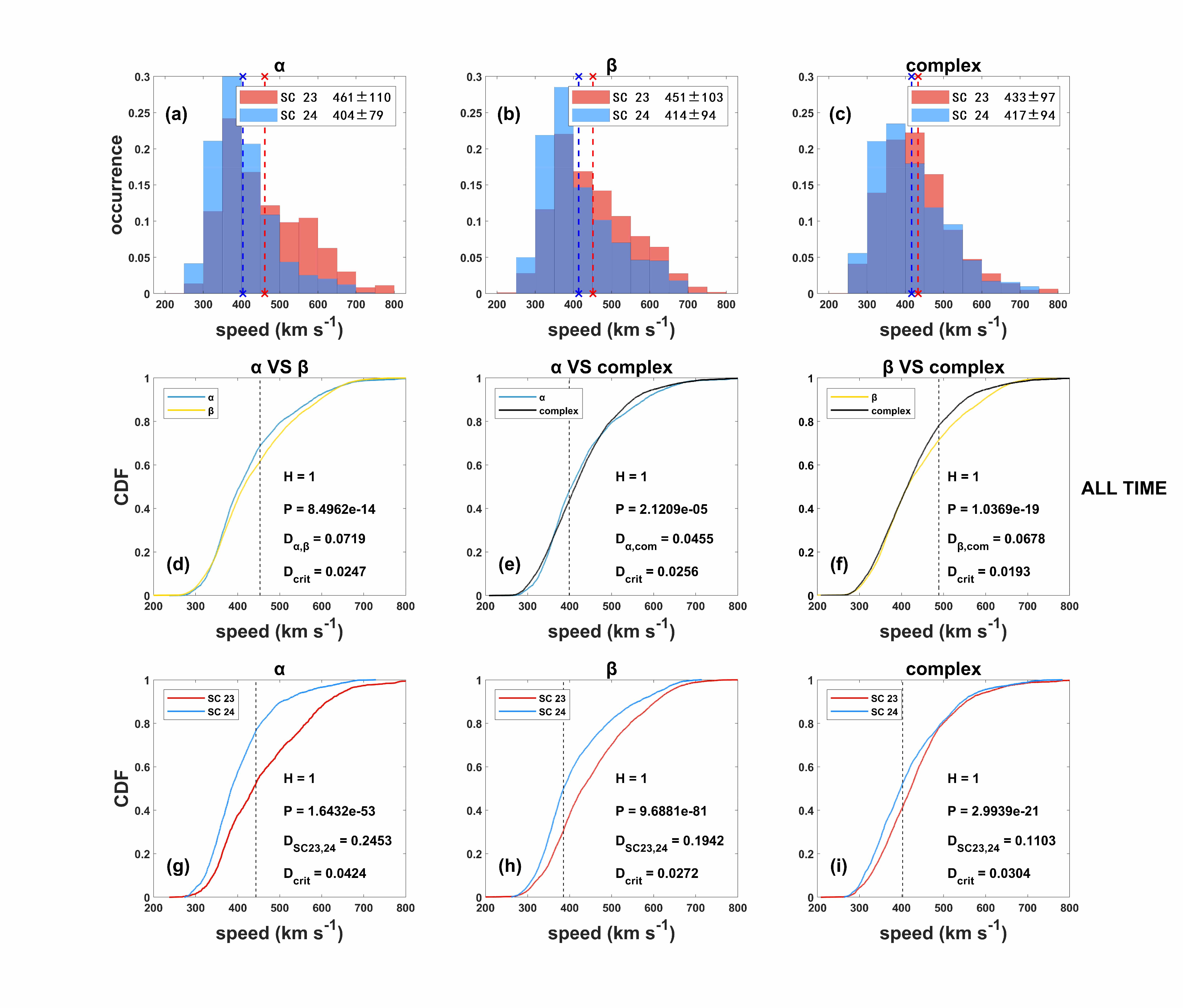}
\caption{The speeds of $\alpha$, $\beta$, and complex AR solar wind. The layout is the same as Figure \ref{footpoint}. The speeds of three types of AR solar wind are generally the same.}
\label{speed}
\end{figure}
Figure \ref{speed} presents the speeds of these three types of AR solar wind. The speed histograms of the three types of AR solar wind during SCs 23 (red) and 24 (blue) are shown in the panels (a)-(c) of Figure \ref{speed}. The average speeds for $\alpha$, $\beta$, and complex AR solar wind during SC 23 (24) are 461 (404) \velunit, 451 (414) \velunit, and 433 (417) \velunit, respectively. The speeds of the three types of AR solar wind are generally similar. The CDF curves for the three types of AR solar wind are also quite similar, although the two sample K-S tests show differences in their speed distributions (see the panels (d)-(f) of Figure \ref{speed}). All D values are greater than their corresponding $D_{\mathrm{crit}}$ values, and all P values are smaller than 0.05. Therefore, all H values are 1. For the same type of AR solar wind, the speeds during SC 23 are generally higher than those during SC 24, especially for $\alpha$ and $\beta$ AR solar wind (see panels (g)-(i) in Figure \ref{speed}).

\subsection{The $O^{7+}/O^{6+}$ and $Q_{Fe}$ in the three types of AR solar wind during SCs 23 and 24}
\begin{figure}[!h]
\vspace{-1.5cm}
\hspace{-2cm}
\setlength{\abovecaptionskip}{-1cm}
\includegraphics[width=1.2\textwidth]{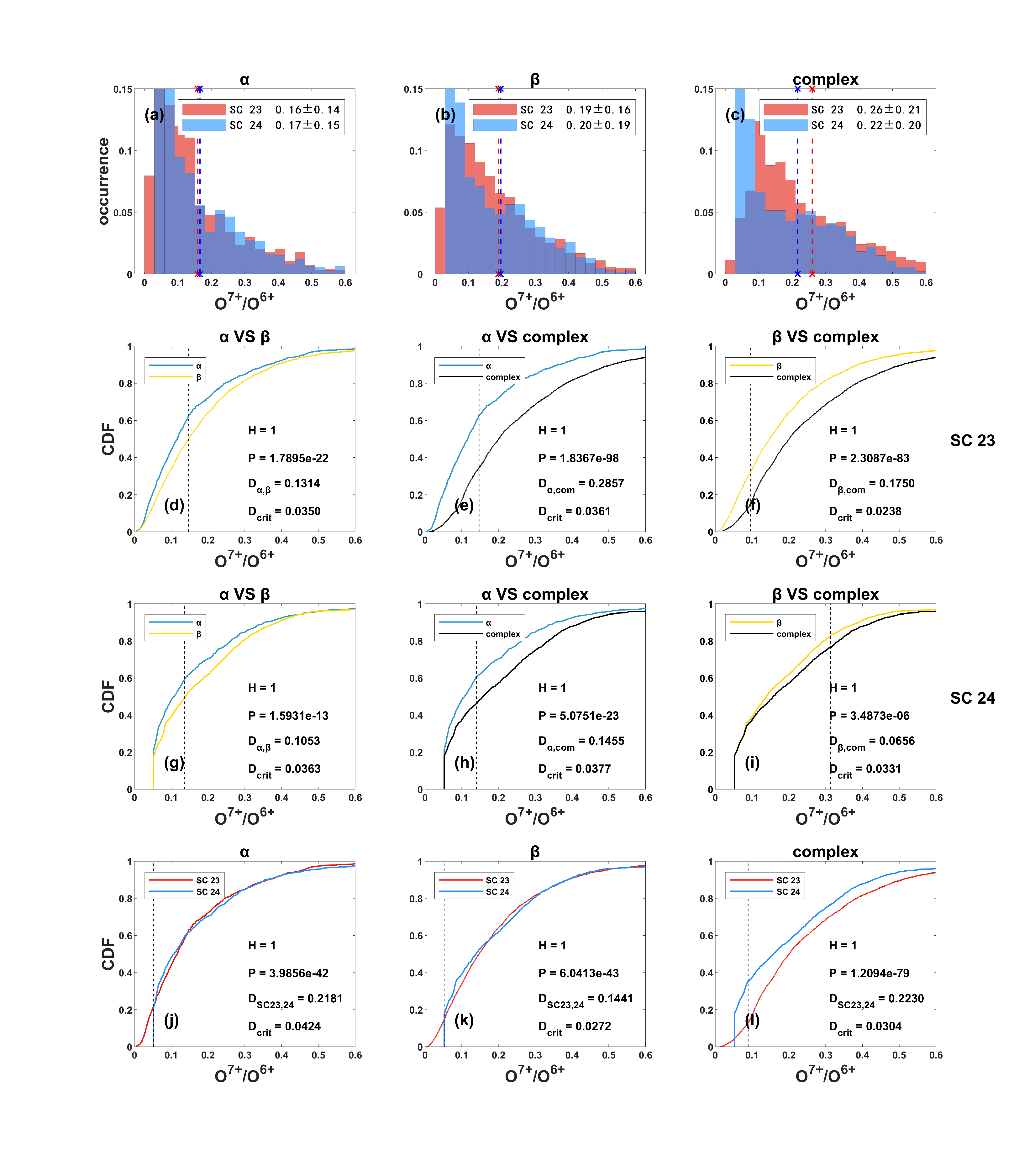}
\caption{The $O^{7+}/O^{6+}$ of $\alpha$, $\beta$, and complex AR solar wind during SCs 23 and 24. The first row shows the histograms of $O^{7+}/O^{6+}$ of the three types of AR solar wind during SC 23 (red) and SC 24 (blue). The second and third rows present the CDF curves of $O^{7+}/O^{6+}$ for different types of AR solar wind during SC 23 and SC 24, respectively. The cyan, yellow, and black curves represent $\alpha$, $\beta$, and complex AR solar wind, respectively. The fourth row compares the CDF curves of $O^{7+}/O^{6+}$ for $\alpha$, $\beta$, and complex AR solar wind between SC 23 and SC 24. The difference of $O^{7+}/O^{6+}$ in the three types of AR solar wind is obvious.}
\label{O76}
\end{figure}
\begin{figure}[!h]
\hspace{-2cm}
\setlength{\abovecaptionskip}{-1cm}
\includegraphics[width=1.2\textwidth]{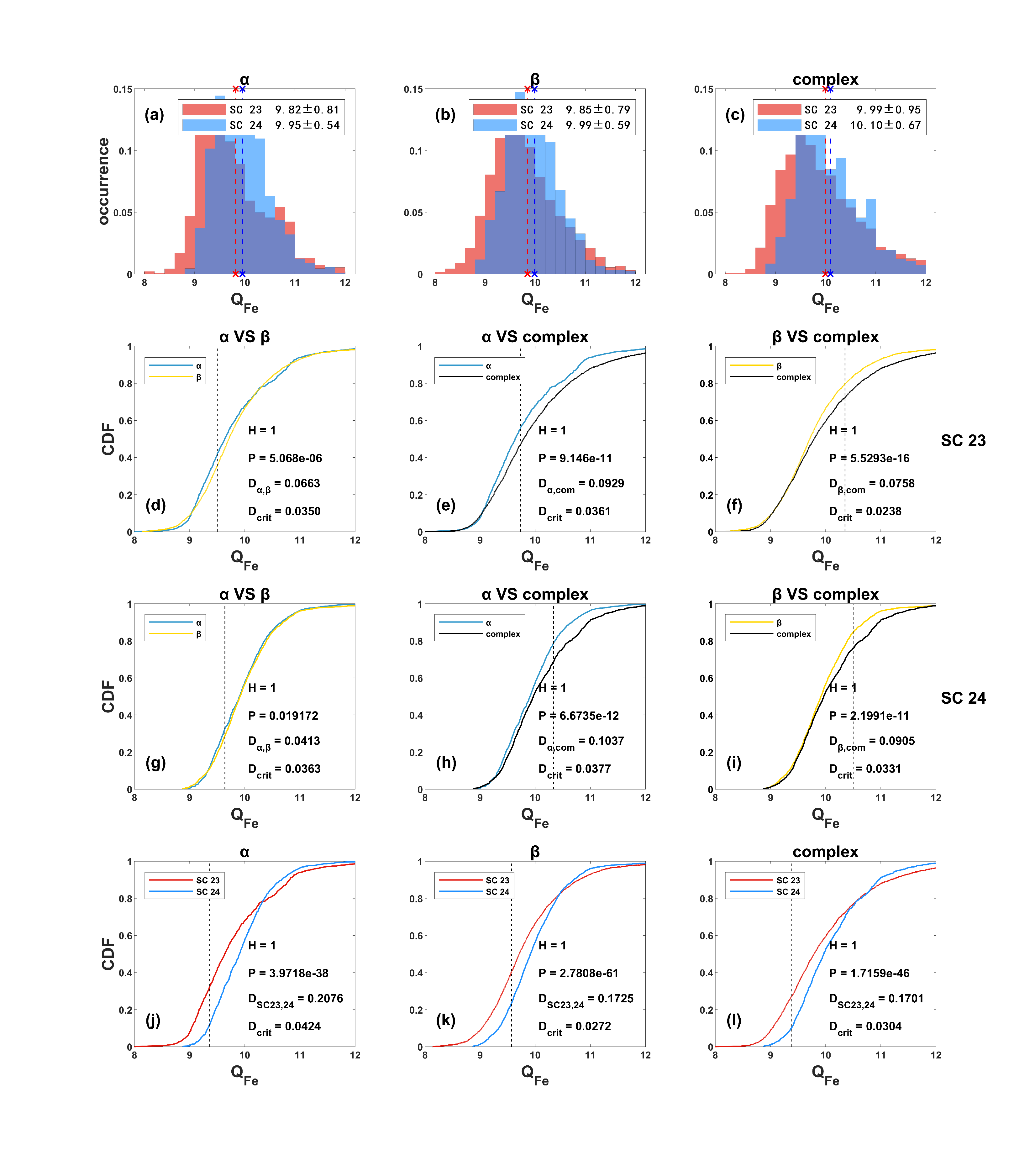}
\caption{The $Q_{Fe}$ in the three types of AR solar wind. The layout is the same as Figure \ref{O76}. As the magnetic complexity of the AR increases, the $Q_{Fe}$ in the AR solar wind also increases.}
\label{QFe}
\end{figure}

The density ratio of $O^{7+}$ and $O^{6+}$ ($O^{7+}/O^{6+}$) in the three types of AR solar wind are shown in Figure \ref{O76}. The panels (a)-(c) of Figure \ref{O76} show the statistical results of $O^{7+}/O^{6+}$ in solar wind originating from the three types of ARs during SCs 23 (red) and 24 (blue). The average $O^{7+}/O^{6+}$ in $\alpha$, $\beta$, and complex AR solar wind during SC 23 (24) are 0.16 (0.17), 0.19 (0.20), and 0.26 (0.22), respectively. The statistical results indicate that the $O^{7+}/O^{6+}$ in AR solar wind is strongly influenced by the magnetic field structure in the source region. The higher the magnetic complexity of ARs, the higher the $O^{7+}/O^{6+}$ in the corresponding AR solar wind. The ACE/SWICS data before and after the anomaly are not suitable to be combined. Therefore, for $O^{7+}/O^{6+}$, $Q_{Fe}$, and FIP bias, we perform the two sample K-S test separately for different types of AR solar wind in SC 23 and SC 24 (see panels (d)-(i) in Figures \ref{O76}, \ref{QFe} and \ref{FIP}). The CDF curves and the results of the two sample K-S test further confirm that the magnetic complexity of ARs significantly affects the $O^{7+}/O^{6+}$ in AR solar wind. The results of two sample K-S test show that all D values are significantly greater than their corresponding $D_{\mathrm{crit}}$ values, and all P values are significantly smaller than 0.05. Therefore, all H values equal 1. There are significant differences in the distribution of $O^{7+}/O^{6+}$ among the three types of AR solar wind. The $O^{7+}/O^{6+}$ in the complex AR solar wind is much higher than that in $\alpha$ and $\beta$ AR solar wind. As the magnetic complexity of ARs increases, the difference in $O^{7+}/O^{6+}$ of the same type of AR solar wind between SCs 23 and 24 becomes larger. The $O^{7+}/O^{6+}$ in the complex AR solar wind during SC 23 is much higher than that during SC 24.

The average charge states of iron ($Q_{Fe}$) in the three types of AR solar wind are presented in Figure \ref{QFe}. The histograms of $Q_{Fe}$ in the three types of AR solar wind during SCs 23 (red) and 24 (blue) are presented in the panels (a)-(c) of Figure \ref{QFe}. The average $Q_{Fe}$ in $\alpha$, $\beta$, and complex AR solar wind are 9.82 (9.95), 9.85 (9.99), and 9.99 (10.10) during SC 23 (24), respectively. The $Q_{Fe}$ in AR solar wind is influenced to some extent by the magnetic complexity of ARs. The $Q_{Fe}$ in solar wind from complex ARs is slightly higher. In both SC 23 and SC 24, the CDF curves of $Q_{Fe}$ in $\alpha$ and $\beta$ AR solar wind are close (see the panels (d) and (g) in Figure \ref{QFe}). In contrast, the $Q_{Fe}$ in complex AR solar wind shows larger differences compared to $\alpha$ and $\beta$ AR solar wind. Similar to $O^{7+}/O^{6+}$, the results of the two sample K-S test also show that all D values are significantly greater than their corresponding $D_{\mathrm{crit}}$ values, and all P values are much smaller than 0.05. Therefore, all H values are 1. The CDF curves of $Q_{Fe}$ for complex AR solar wind begin to diverge significantly from those of $\alpha$ and $\beta$ AR solar wind at $Q_{Fe}$ greater than about 10, suggesting that solar wind from complex ARs has higher $Q_{Fe}$. This indicates that complex ARs are more effective in heating plasma in the solar wind. For the same type of AR solar wind, the histograms show that the $Q_{Fe}$ values in SC 24 are slightly higher than those in SC 23 (see panels (a)-(c) in Figure \ref{QFe}). However, a comparison of the CDF curves of $Q_{Fe}$ for the same AR solar wind types between SC 23 and SC 24 reveals that when $Q_{Fe}$ is below approximately 10.5, the values in SC 24 are higher; whereas when $Q_{Fe}$ exceeds about 10.5, the values in SC 24 become lower than those in SC 23 (see panels (j)-(l) in Figure \ref{QFe}).

The charge states of solar wind can be used to characterize the temperature of the source region. The complexity of the magnetic structure in the source region is typically associated with more frequent magnetic activity, such as the emergence of new magnetic flux, interactions between different magnetic flux systems, and magnetic reconnection. \cite{2016ApJ...820L..11J} pointed out that complex ARs are caused by collisions between different magnetic systems when magnetic flux emerges frequently. Such processes may lead to enhanced localized heating, thereby elevating the plasma temperature in the source region. As a result, with increasing magnetic complexity of the source region, the proportion of high charge-state ions in AR solar wind also increases. Previous studies have shown that the freezing height of oxygen ions is around 0.5 R$_{\bigodot}$, while that of iron ions is about 3-5 R$_{\bigodot}$ \citep{1997SoPh..171..345K,1998ApJ...498..448E,2001ApJ...563.1055E}. The freezing height of oxygen ions is lower than that of iron ions, meaning that $O^{7+}/O^{6+}$ reflects the temperature at lower altitudes in the source region, while $Q_{Fe}$ represents the temperature at higher altitudes. Consequently, $O^{7+}/O^{6+}$ is more sensitive to the magnetic complexity in the source region. As the magnetic complexity of ARs increases, the $O^{7+}/O^{6+}$ in the solar wind from these regions rises more significantly. For solar wind from complex ARs, the $O^{7+}/O^{6+}$ is significantly higher compared to that in the $\alpha$ and $\beta$ AR solar wind. In contrast, because the freezing height of iron ions is higher, the $Q_{Fe}$ in the solar wind from ARs is less affected by the magnetic activity of the source region. As a result, the $Q_{Fe}$ in solar wind from $\alpha$ and $\beta$ ARs is relatively similar, with only the $Q_{Fe}$ in complex AR solar wind being slightly higher than that in $\alpha$ and $\beta$ AR solar wind. Our results show that, both in SC 23 (before the ACE/SWICS anomaly) and SC 24 (after the ACE/SWICS anomaly), the increase in $O^{7+}/O^{6+}$ with increasing source region magnetic complexity is more pronounced than that of $Q_{Fe}$ in AR solar wind. In SC 23 (SC 24), the average $O^{7+}/O^{6+}$ for $\alpha$, $\beta$, and complex AR solar wind are 0.16 (0.17), 0.19 (0.20), and 0.26 (0.22), respectively; while the average $Q_{Fe}$ are 9.82 (9.95), 9.85 (9.99), and 9.99 (10.10), respectively.

Due to a hardware anomaly on August 23, 2011, ACE/SWICS changes its operating mode, resulting in inconsistent calibration standards before and after the anomaly. Therefore, care must be taken not to mix ACE/SWICS data from before and after the anomaly. Our current results show that in both SC 23 and SC 24, the charge states of AR solar wind increase with increasing magnetic complexity of the source region. This conclusion is not affected by the ACE/SWICS anomaly. However, differences in charge states between the same type of AR solar wind in the two solar cycles may be influenced both by the solar cycle amplitude and by the instrument anomaly. Due to the interplay between the two factors, it is difficult to determine which exerts the primary influence. Nevertheless, the effect of the instrument anomaly is unequivocally present. According to the ACE/SWICS 2.0 data release notes, saturation issues in SWICS 2.0 most commonly affect $O^{7+}/O^{6+}$, while $N_{Fe}/N_{O}$ is less affected, and $Q_{Fe}$ is least affected. During the period 2008-2010, the actual $O^{7+}/O^{6+}$ in fast solar wind is often several times lower than the value given by SWICS 2.0. This indicates that SWICS 2.0 may systematically overestimate $O^{7+}/O^{6+}$, and our results support this. The $O^{7+}/O^{6+}$ in AR solar wind during SC 24 (after the anomaly) shows a cutoff at 0.0523, with no values falling below this threshold. Previous studies show that SC 23 has higher sunspot numbers and overall solar activity than SC 24. Therefore, AR solar wind in SC 23 is expected to exhibit higher $O^{7+}/O^{6+}$. However, for $\alpha$ and $\beta$ AR solar wind, the $O^{7+}/O^{6+}$ in SC 24 are slightly higher than in SC 23. In contrast, complex AR solar wind shows significant differences in the $O^{7+}/O^{6+}$ distribution between SC 23 and SC 24, with values in SC 23 being clearly higher than in SC 24. This implies that SWICS 2.0 may systematically overestimate $O^{7+}/O^{6+}$, suggesting that the actual $O^{7+}/O^{6+}$ of AR solar wind may have been lower after the anomaly. We suspect that this bias is more pronounced in ARs with lower magnetic complexity (such as $\alpha$ and $\beta$ ARs). Although higher overall solar activity in SC 23 should result in higher $O^{7+}/O^{6+}$, we find that the $O^{7+}/O^{6+}$ for $\alpha$ and $\beta$ AR solar wind is slightly higher in SC 24, which may be attributed to the effects of the instrument anomaly. However, it is worth noting that in complex AR solar wind, $O^{7+}/O^{6+}$ in SC 23 remains significantly higher than in SC 24. This may imply that when the source region has more complex magnetic fields and the solar wind is more intensely heated, the saturation effect in $O^{7+}/O^{6+}$ due to the SWICS anomaly may be alleviated.
As for $Q_{Fe}$ in AR solar wind, SWICS 2.0 may also overestimate values after the anomaly. Observations show that for all three types of AR solar wind, $Q_{Fe}$ in SC 23 (with stronger solar activity) is lower than in SC 24 (with weaker solar activity). This may be related to the SWICS instrument anomaly. The CDF curves show this trend clearly (the fourth row of Figure \ref{QFe}). When $Q_{Fe} < 10.5$, SC 24 are higher than SC 23. When $Q_{Fe} > 10.5$, SC 24 becomes lower than SC 23. This pattern may indicate a systematic overestimation of $Q_{Fe}$ in the SWICS 2.0 dataset, especially in weaker source regions. According to current understanding, the overall solar activity in SC 23 is significantly stronger than in SC 24. Therefore, the AR solar wind in SC 23 is theoretically expected to have higher $Q_{Fe}$. However, observations show that $Q_{Fe}$ in all three types of AR solar wind is actually lower in SC 23 than in SC 24. This supports the possibility that SWICS 2.0 overestimates $Q_{Fe}$ after the anomaly. This effect is more evident in source regions with lower magnetic complexity. When $Q_{Fe}$ is low, values in SC 23 are lower than in SC 24. When $Q_{Fe}$ is high, values in SC 23 are higher than in SC 24. This suggests that when the source region experiences strong heating (with higher $Q_{Fe}$), the overestimation effect of $Q_{Fe}$ by SWICS 2.0 may be reduced. Based on the analysis of $O^{7+}/O^{6+}$ and $Q_{Fe}$, it appears that after the SWICS instrument anomaly, the charge states of AR solar wind may be significantly overestimated when the magnetic complexity of the source region is low. As the magnetic complexity increases, this issue may become less pronounced.

\subsection{The $A_{He}$ and FIP bias in the three types of AR solar wind during SCs 23 and 24}
\begin{figure}[!h]
\hspace{-2cm}
\setlength{\abovecaptionskip}{-0.5cm}
\includegraphics[width=1.2\textwidth]{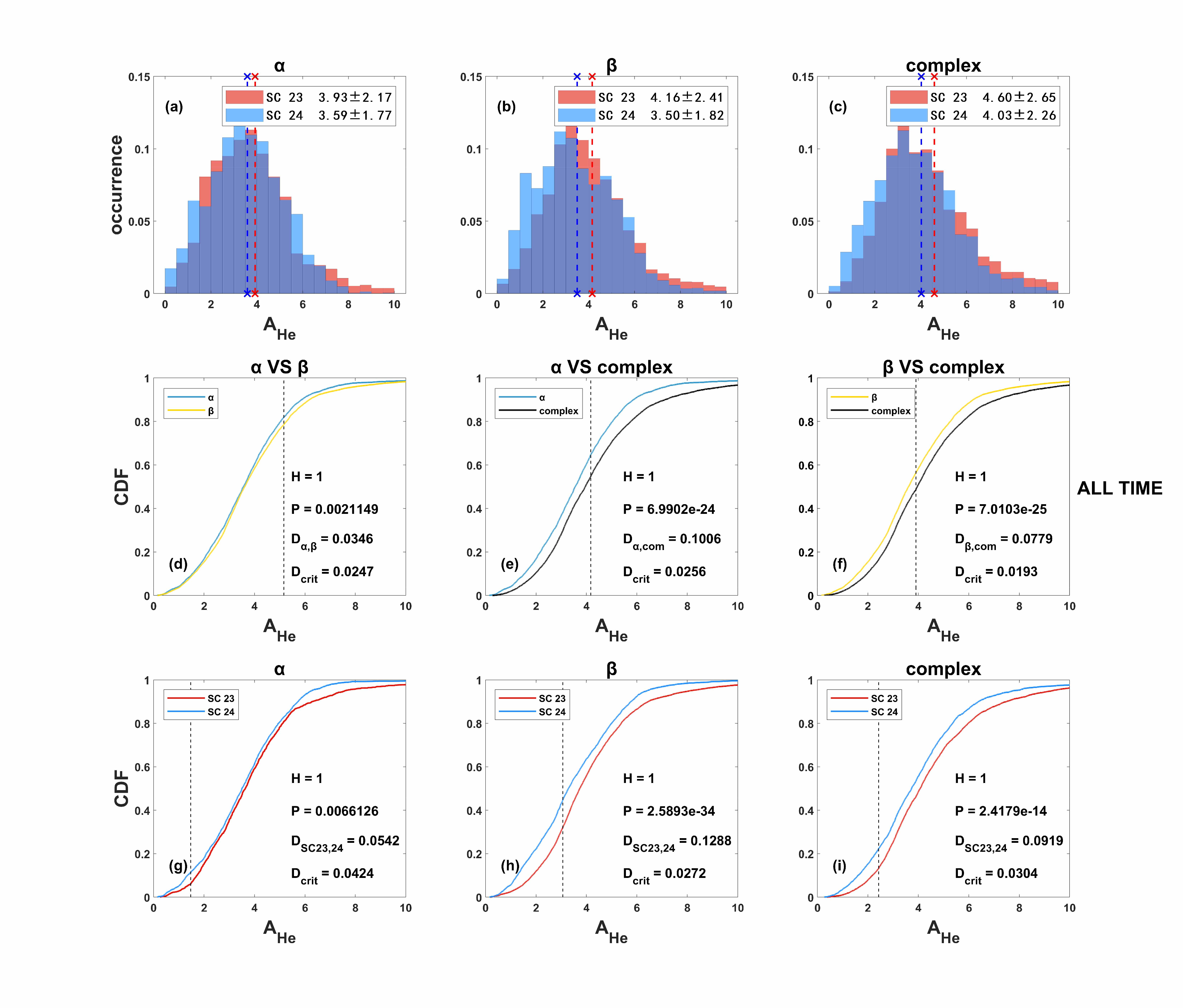}
\caption{The $A_{He}$ in three types of AR solar wind. The layout is the same as Figure \ref{footpoint}. The $A_{He}$ in complex AR solar wind is significantly higher than that in $\alpha$ and $\beta$ AR solar wind.}
\label{AHe}
\end{figure}
The $A_{He}$ in three types of AR solar wind are shown in Figure \ref{AHe}. The panels (a)-(c) of Figure \ref{AHe} show the histograms of $A_{He}$ in solar wind from different types of ARs during SCs 23 (red) and 24 (blue). The average $A_{He}$ in $\alpha$, $\beta$, and complex AR solar wind during SC 23 (24) are 3.93 (3.59), 4.16 (3.50), and 4.60 (4.03), respectively.
Overall, similar to other parameters of AR solar wind presented above, $A_{He}$ in AR solar wind is also influenced by the magnetic structure of the source region. Solar wind originating from complex ARs has significantly higher $A_{He}$ compared to solar wind from $\alpha$ and $\beta$ ARs. The $A_{He}$ distributions of $\alpha$ and $\beta$ AR solar wind are very similar, as their CDF curves almost overlap (see the panel (d) of Figure \ref{AHe}). In contrast, solar wind from complex ARs shows noticeable differences in $A_{He}$ compared to the solar wind from $\alpha$ and $\beta$ ARs. This result suggests that complex ARs are more able to transport the helium-rich materials into the solar wind. The two sample K-S test results still show that all D values are significantly greater than the $D_{\mathrm{crit}}$ values and all P values are significantly smaller than 0.05; therefore, all H values are 1. Additionally, for the same type of AR solar wind, there are differences in $A_{He}$ between the two solar cycles. AR solar wind during SC 23 generally has higher $A_{He}$ than that during SC 24, and this difference increases as the magnetic complexity of the AR increases. This is consistent with previous studies, as the $A_{He}$ in AR solar wind is also influenced by the magnetic activity of the source region and the amplitude of the solar cycle \citep{2019ApJ...879L...6A,2021SoPh..296...67A,2021MNRAS.503L..17Y,2022ApJ...925..137S,2024ApJ...970L..16O}. In addition, \cite{2025ApJ...982L..40A} and \cite{2025A&A...694A.265A} pointed out that $A_{He}$ in the solar wind exhibits two distinct gradient trends with speeds above and below 400 \velunit. The result of \cite{2025ApJ...982L..40A} shows that in the slow solar wind, $A_{He}$ increases monotonically from 0\% to 4.19\%, while in the fast solar wind (speed $>$ 433 \velunit), it tends to saturate at $A_{He}$ = 4.19\%. Our results show that the speed and $A_{He}$ values of the AR solar wind are consistent with those of the slow wind studied by \cite{2025ApJ...982L..40A} and \cite{2025A&A...694A.265A}. This may suggest that at least part of the slow solar wind observed in their studies originates from ARs.

\begin{figure}[!h]
\hspace{-2cm}
\setlength{\abovecaptionskip}{-1cm}
\includegraphics[width=1.2\textwidth]{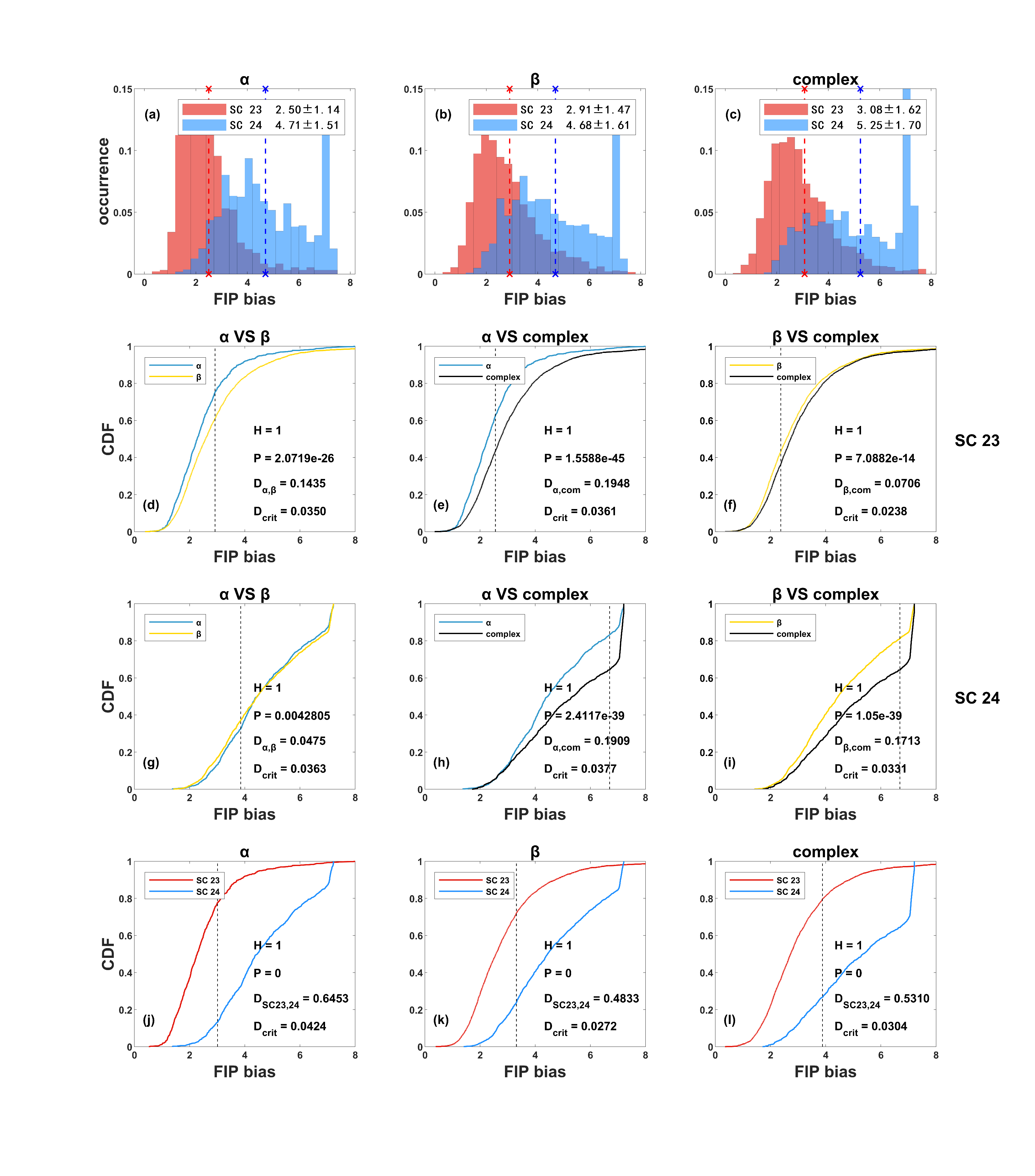}
\caption{The FIP bias of the three types of AR solar wind. The layout is the same as Figure \ref{O76}. The FIP bias is increased with the complexity of the source region magnetic field structure. }
\label{FIP}
\end{figure}
The FIP bias of the three types of AR solar wind are given in Figure \ref{FIP}. The FIP bias describes the phenomenon of low-FIP elements, such as Mg, Si, and Fe, being enriched in the corona and solar wind \citep{1963ApJ...137..945P,2015LRSP...12....2L}. The FIP bias can be determined using ($N_{Fe}/N_{O}$)/0.06 \citep{2017ApJ...836..169F}. The panels (a)-(c) of Figure \ref{FIP} shows the statistical results of FIP bias in solar wind from three types of ARs during SC 23 and SC 24. The average FIP bias in solar wind from $\alpha$, $\beta$, and complex ARs during SC 23 (24) are 2.50 (4.71), 2.91 (4.68), and 3.08 (5.25), respectively. Similar to $A_{He}$, the FIP bias of AR solar wind is also influenced by the magnetic complexity of ARs, the FIP bias of the complex AR solar wind is the highest. The results of the CDF curves show that, in both SC 23 and SC 24, the CDF curve of the FIP bias for complex AR solar wind lies below those of the $\alpha$ and $\beta$ AR solar wind. All D values are significantly greater than the $D_{\mathrm{crit}}$ values and all P values are significantly smaller than 0.05; therefore, all H values are 1 (see the panels (d)-(i) of Figure \ref{FIP}). The results of the two sample K-S test also indicate that the FIP bias distributions differ among different types of AR solar wind.

Similar to $O^{7+}/O^{6+}$ and $Q_{Fe}$, the FIP bias is also affected by the ACE/SWICS anomaly. From the histogram and CDF curves (first and fourth rows of Figure \ref{FIP}), the FIP bias distributions of the same types of AR solar wind show clear differences between SC 23 and SC 24, the FIP bias in SC 24 is significantly higher than in SC 23, suggesting that the SWICS instrument anomaly also significantly affects FIP bias. In SC 24, the FIP bias distributions of AR solar wind shift to higher values and show saturation in the high-value range. This may also indicate a systematic overestimation of FIP bias in SWICS 2.0. The two sample K-S tests confirm that the differences in FIP bias distributions of same types of AR solar wind between SC 23 and SC 24 are statistically significant. The FIP bias differences between SCs 23 and 24 must be due to anomalies in the ACE/SWICS instrument. Similar to the charge states, the FIP bias measurements after the instrumental anomaly may be significantly overestimated. In summary, the SWICS instrument anomaly significantly affects the measurements of $O^{7+}/O^{6+}$, $Q_{Fe}$, and FIP bias, with more pronounced impacts on $O^{7+}/O^{6+}$ and FIP bias. Current results show that under weak magnetic activity in the source region, SWICS 2.0 may systematically overestimate $O^{7+}/O^{6+}$, $Q_{Fe}$, and FIP bias. As magnetic complexity increases, this overestimation may tends to weaken. Although the pre- and post-anomaly data are not suitable for direct combination in long-term trend analyses, they can still be analyzed separately to investigate the effect of source region magnetic activity on AR solar wind properties. This approach is valid because the calibration methods used before and after the SWICS anomaly remain consistent, ensuring internal consistency within each dataset. Moreover, at both stages, the distributions of $O^{7+}/O^{6+}$, $Q_{Fe}$, and FIP bias among different AR solar wind types show significant differences. All three parameters of AR solar wind increase overall with the complexity of the source regions.

The $A_{He}$ distributions of $\alpha$ and $\beta$ AR solar wind are similar, while solar wind from complex ARs shows significantly higher $A_{He}$. A similar trend is observed for FIP bias. As the magnetic complexity of the source region increases, the FIP bias of AR solar wind also increases. The source and composition of plasma within ICMEs are highly complex and influenced by various physical mechanisms, including chromospheric evaporation, gravitational settling, and the FIP effect \citep{2020ApJ...900L..18F,2021ApJ...912...51L,2022ApJ...936...83R,2022MNRAS.513L.106Y}. Previous studies show that the increased $A_{He}$ in ICMEs is closely related to chromospheric activity at the base of the source region \citep{2020ApJ...900L..18F,2022MNRAS.513L.106Y}. \cite{2020ApJ...900L..18F} found that during flares, helium-rich materials from the chromosphere can be transported into the ICME through chromospheric evaporation processes. Statistical results show that as flare intensity increases, the proportion of helium-rich materials from chromospheric evaporation processes in the ICMEs also increases significantly \citep{2023ApJ...956..129F}. These findings indicate that the $A_{He}$ within ICMEs is closely linked to the magnetic activity of their source regions. Similarly, numerous studies have shown that the $A_{He}$ in the solar wind is also closely related to the magnetic structure and activity level of its source region \citep{2024ApJ...977...89Y,2025A&A...694A.265A,2025ApJ...982L..40A}. Previous research indicates that source regions with persistently open magnetic field structures tend to produce solar wind with higher and more stable $A_{He}$, while solar wind originating from regions with intermittently open magnetic fields exhibits lower and more dynamically variable $A_{He}$. This suggests that the magnetic topology and activity of the source region are likely to play an important role in influencing the $A_{He}$ in both ICMEs and the solar wind. So the above helium-rich materials supply scenario for ICMEs may also be valid for the AR solar wind. Therefore, we hypothesize that the $A_{He}$ in the AR solar wind may also be closely linked to the magnetic activity of its source region. More complex magnetic structures are associated with more frequent magnetic activities, leading to higher $A_{He}$ in AR solar wind. \cite{2016ApJ...820L..11J} pointed out that complex ARs are caused by frequent magnetic flux emergence and the interactions between different magnetic flux systems. ARs with complex magnetic structures should be associated with frequent mini-flares and chromospheric evaporation processes. These processes may transport helium-rich materials from the chromosphere into the upper solar atmosphere and the solar wind. Therefore, in complex ARs, increased magnetic reconnection and convective processes may effectively transport helium-rich materials from the lower atmosphere to the upper corona, thus influencing the solar wind composition. Frequent magnetic activity also intensifies the heating in the source region, resulting in the solar wind from complex ARs that has significantly higher charge states and $A_{He}$ compared to the solar wind from $\alpha$ and $\beta$ ARs. This also explains why the $A_{He}$ in the slow solar wind is correlated with sunspot numbers \citep{2001GeoRL..28.2767A,2007ApJ...660..901K,2012ApJ...745..162K,2019ApJ...879L...6A,2021SoPh..296...67A,2022ApJ...925..137S}. ACE and WIND are both located in the ecliptic plane and primarily detect solar wind originating from low and mid-latitudes. Therefore, they are capable of effectively observing AR solar wind from these latitudes throughout all phases of the solar cycle. Our results show that during solar maximum, the magnetic fields of ARs are more complex and active, which may facilitate the injection of helium-rich material from lower atmospheric layers into the corona and its subsequent involvement in solar wind formation. This results in a higher $A_{He}$ in AR solar wind. In contrast, during solar minimum, the Sun becomes relatively quiet, reduced magnetic activity in ARs leads to a lower $A_{He}$. Therefore, we conclude that the solar cycle variation in $A_{He}$ in AR solar wind is mainly influenced by the magnetic activity of active regions.


Previous observational studies have shown that regions with strong magnetic fields exhibit the highest FIP bias \citep{2013ApJ...778...69B}. The FIP bias statistical results of the AR solar wind in our study are consistent with the remote observations. In ARs with the strongest and most complex magnetic fields, the solar wind exhibits the highest FIP bias. This result indicates that as the magnetic complexity of the source region increases, the proportion of plasma originating from the lower chromosphere injected into the corona and solar wind also increases. This plasma likely resides in a temperature range of 10,000 to 50,000 K before being heated. At these temperatures, helium remains neutral while low-FIP elements such as magnesium, silicon, and iron are already ionized. The helium is not depleted and the FIP bias effect is valid (low FIP elements are enriched) for the materials in the above temperatures.
Therefore, the above materials should be associated with higher helium abundance and FIP bias effect. Frequent magnetic flux emergence and reconnection in complex ARs transport this material to the upper corona, from where it is released, resulting in increased $A_{He}$ and FIP bias in the solar wind from complex ARs.

\section{Summary and Conclusions} \label{sec:displaymath}
\begin{table}[!htbp]
\scriptsize
\centering
\caption{The mean, standard deviation (SD), and standard error of the mean (SEM) for the properties of the three types of AR solar wind in SC 23 and SC 24.}
\label{tab3}
\begin{tabular}{ccccccccccc}
\midrule
\midrule
 \multirow{2}{*}{Parameter}                          &         & \multicolumn{3}{c}{SC 23} & \multicolumn{3}{c}{SC 24} & \multicolumn{3}{c}{ALL TIME} \\
\cmidrule(r){3-5} \cmidrule(r){6-8} \cmidrule(r){9-11}
                          &         & Mean   & SD     & SEM     & Mean    & SD    & SEM     & Mean   & SD       & SEM      \\
\cmidrule(r){3-5} \cmidrule(r){6-8} \cmidrule(r){9-11}
\midrule
\multirow{3}{*}{footpoint magnetic field strength (G)}     & $\alpha$       & 114    & 141    & 3.2897  & 93      & 121   & 2.6501  & 103    & 131      & 2.0942   \\
                          & $\beta$       & 141    & 126    & 1.4865  & 96      & 95    & 1.5959  & 126    & 118      & 1.1450   \\
                          & complex & 182    & 136    & 1.9051  & 161     & 142   & 2.6925  & 175    & 139      & 1.5604   \\
\midrule
\multirow{3}{*}{in-situ magnetic field strength (nT)}     & $\alpha$       & 5.50   & 2.94   & 0.0683  & 4.87    & 2.02  & 0.0429  & 5.15   & 2.50     & 0.0392   \\
                          & $\beta$       & 6.04   & 5.11   & 0.0597  & 5.19    & 2.35  & 0.0389  & 5.76   & 4.40     & 0.0420   \\
                          & complex & 6.32   & 3.21   & 0.0430  & 5.31    & 2.44  & 0.0445  & 5.96   & 3.00     & 0.0324   \\
\midrule
\multirow{3}{*}{speed (\velunit)}       & $\alpha$       & 461    & 110    & 2.5557  & 404     & 79    & 1.6777  & 430    & 98.50    & 1.5451   \\
                          & $\beta$       & 451    & 103    & 1.2074  & 414     & 94    & 1.5562  & 439    & 101.66   & 0.9717   \\
                          & complex & 433    & 97     & 1.2987  & 417     & 94    & 1.7035  & 427    & 96.07    & 1.0368   \\
\midrule
\multirow{3}{*}{$O^{7+}/O^{6+}$}      & $\alpha$       & 0.16   & 0.14   & 0.0033  & 0.17    & 0.15  & 0.0031  & $\cdots$      & $\cdots$        & $\cdots$        \\
                          & $\beta$       & 0.19   & 0.16   & 0.0019  & 0.20    & 0.19  & 0.0033  & $\cdots$      & $\cdots$        & $\cdots$        \\
                          & complex & 0.26   & 0.21   & 0.0028  & 0.22    & 0.20  & 0.0039  & $\cdots$      & $\cdots$        & $\cdots$        \\
\midrule
\multirow{3}{*}{$Q_{Fe}$}      & $\alpha$       & 9.82   & 0.81   & 0.0189  & 9.95    & 0.54  & 0.0114  & $\cdots$      & $\cdots$        & $\cdots$        \\
                          & $\beta$       & 9.85   & 0.79   & 0.0094  & 9.99    & 0.59  & 0.0100  & $\cdots$      & $\cdots$        & $\cdots$        \\
                          & complex & 9.99   & 0.95   & 0.0128  & 10.10   & 0.67  & 0.0127  & $\cdots$      & $\cdots$        & $\cdots$        \\
\midrule
\multirow{3}{*}{$A_{He}$}      & $\alpha$       & 3.93   & 2.17   & 0.0520  & 3.59    & 1.77  & 0.0382  & 3.74   & 1.97     & 0.0315   \\
                          & $\beta$       & 4.16   & 2.41   & 0.0290  & 3.50    & 1.82  & 0.0305  & 3.93   & 2.25     & 0.0220   \\
                          & complex & 4.60   & 2.65   & 0.0361  & 4.03    & 2.26  & 0.0420  & 4.40   & 2.53     & 0.0278   \\
\midrule
\multirow{3}{*}{FIP bias} & $\alpha$       & 2.50   & 1.14   & 0.0268  & 4.71    & 1.51  & 0.0323  & $\cdots$      & $\cdots$        & $\cdots$        \\
                          & $\beta$       & 2.91   & 1.47   & 0.0174  & 4.68    & 1.61  & 0.0272  & $\cdots$      & $\cdots$        & $\cdots$        \\
                          & complex & 3.08   & 1.62   & 0.0220  & 5.25    & 1.70  & 0.0324  & $\cdots$      & $\cdots$        & $\cdots$        \\
\midrule
\end{tabular}
\end{table}
In the present study, we classify ARs based on the Mount Wilson magnetic classification. Then the AR solar wind from 1999 to 2020 is divided into three categories: $\alpha$, $\beta$, and complex AR solar wind. We statistically analyze the impact of different magnetic types of ARs on solar wind properties. The main findings are as follows:
\begin{enumerate}

\item
The numbers and proportions of the three types of ARs and associated near-Earth AR solar wind are both influenced by solar activity. The near-Earth AR solar wind mainly originates from $\beta$ and complex ARs. Complex ARs  contribute substantially to the AR solar wind, even though they account for a small proportion of all ARs. The proportions of $\alpha$, $\beta$, and complex ARs are 19.99\%, 66.67\%, and 13.34\%, respectively. The percentages of the three types of AR solar wind are 16.96\%, 45.18\%, and 37.86\%, respectively. The proportion of $\alpha$ ($\beta$) AR solar wind during SC 23 is much lower (higher) than that during SC 24. In contrast, the proportion of complex AR solar wind remains nearly the same across the both solar cycles, accounting for about 40\%.

\item
 The properties of AR solar wind are strongly influenced by the magnetic complexity of its source region. The parameters of the three type of AR solar wind in SC 23 and 24 are summarized in Table \ref{tab3}. For the quantities unaffected by the SWICS anomaly, the averages over the entire study period are also included. The more complex the magnetic structure of the AR, the higher the corresponding AR solar wind parameters. The footpoint magnetic field strengths, in-situ magnetic field strengths, charge states ($Q_{Fe}$ and $O^{7+}/O^{6+}$), $A_{He}$, and FIP bias of solar wind from complex ARs are significantly higher than those from $\alpha$ and $\beta$ ARs. The results of the CDF and two sample K-S tests show that there are clear differences in the parameter distributions between AR solar wind from complex ARs and those from $\alpha$ and $\beta$ ARs.

\end{enumerate}

Our statistical results show that complex ARs are more effective at generating solar wind. Additionally, the properties of AR solar wind are strongly influenced by the magnetic structure of its source region. As the magnetic complexity of ARs increases, the footpoint magnetic field strengths, in-situ magnetic field strengths, charge states ($Q_{Fe}$ and $O^{7+}/O^{6+}$), $A_{He}$ and FIP bias of the solar wind also increase. These findings demonstrate that the strong magnetic fields and frequent magnetic activities in complex ARs heat the plasma to higher temperatures and effectively transport helium-rich materials from the lower atmosphere to the upper corona.

\section*{acknowledgments}
\begin{acknowledgments}
The authors thank the anonymous referee for the helpful and constructive comments and suggestions. We appreciate NOAA and the USAF for providing the Solar Region Summary (SRS). In addition, we thank the ACE SWICS instrument team and the ACE Science Center for providing the ACE data. Analysis of Wind SWE observations is supported by NASA grant NNX09AU35G. The yearly sunspot number is provided by the Solar Influence Data Center of the Royal Observatory of Belgium. This research is supported by the National Natural Science Foundation of China (12473058 and 42230203).
\end{acknowledgments}

\bibliography{references1}{}
\bibliographystyle{aasjournal}

\begin{appendix}
\counterwithin{figure}{section}
\counterwithin{table}{section}
\section{Comparison of $O^{7+}/O^{6+}$, $Q_{Fe}$, and FIP bias for all AR wind before and after the ACE/SWICS anomaly.}
\label{sec:appA}
    \begin{figure}[!htbp]
    \hspace{-2cm}
    \centering
    \includegraphics[width=1.1\textwidth]{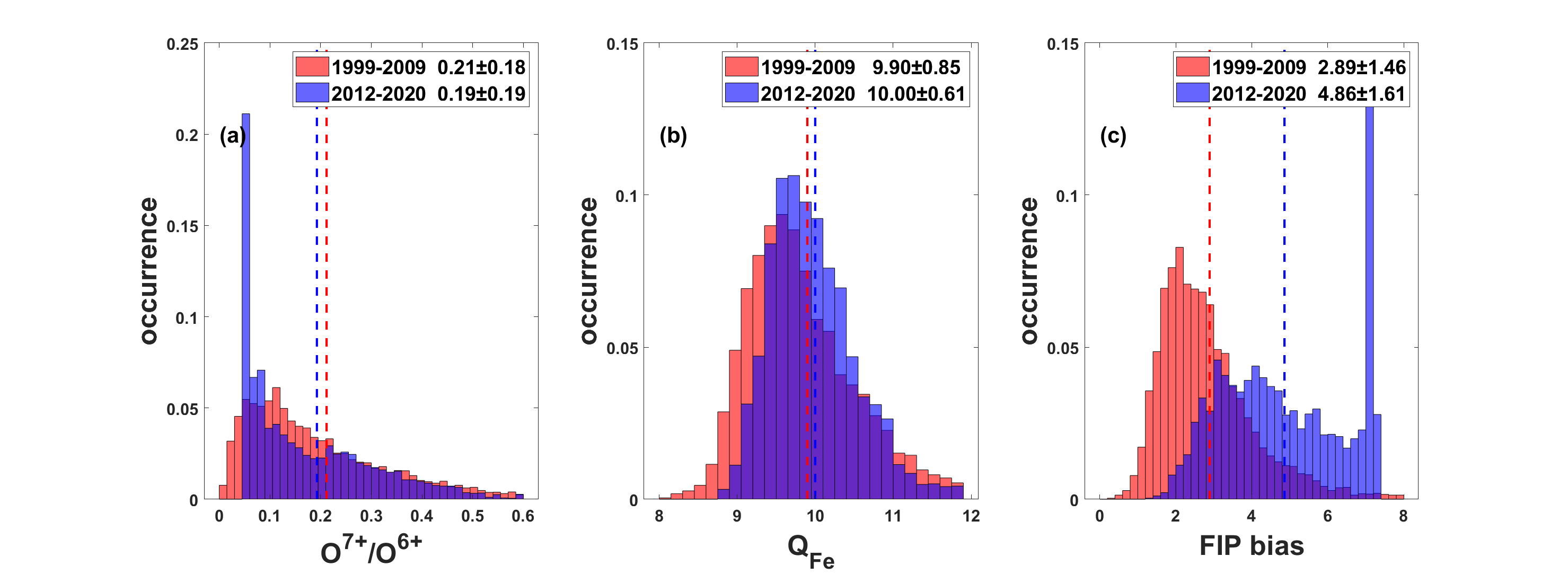}
    \caption{Panels (a), (b), and (c) show the histogram comparisons of $O^{7+}/O^{6+}$, $Q_{Fe}$, and FIP bias for all AR solar wind before and after the ACE/SWICS anomaly, respectively. The dashed lines indicate the corresponding mean values.}
    \label{allwind}
    \end{figure}
The SWICS instrument team has stated that SWICS 1.1 and SWICS 2.0 should not be confused, as they have different measurement and statistical characteristics. In fact, they should be regarded as datasets originating from "different instruments".  SWICS 2.0 may have data saturation issues, which are most commonly observed in $O^{7+}/O^{6+}$, $O^{8+}/O^{6+}$, and $C^{6+}/C^{5+}$. In contrast, $Fe/O$ are less affected, with the $Q_{Fe}$ being the least impacted. To illustrate the impact of the ACE/SWICS anomaly on the data, we compare the statistical results of $O^{7+}/O^{6+}$, $Q_{Fe}$, and FIP bias before and after the anomaly; all the data are from the original observations. The selected time period is consistent with that used in the main text.

Panels (a), (b), and (c) of Figure \ref{allwind} show the histogram comparisons of $O^{7+}/O^{6+}$, $Q_{Fe}$, and FIP bias for all AR solar wind before and after the ACE/SWICS anomaly, respectively. By comparing the histogram distributions, the impact of the ACE/SWICS anomaly on $O^{7+}/O^{6+}$, $Q_{Fe}$, and FIP bias can be clearly observed, with a particularly significant effect on $O^{7+}/O^{6+}$ and FIP bias. For $O^{7+}/O^{6+}$, we find that $O^{7+}/O^{6+}$ exhibits a clear cutoff in the low-value region after the anomaly, and through the original observational data provided by the instrument team, we find that the cutoff value is 0.0523. For FIP bias, the histogram for SC 24 shifts noticeably to the right compared to SC 23. When the FIP bias exceeds 7, the data exhibit a clear saturation effect, and through the original observational data provided by the instrument team, we find that the cutoff is at 7.2117 (see the panel (c) of Figure \ref{allwind}). As for $Q_{Fe}$, the histogram shifts slightly to the right after the anomaly, and no extreme outliers are observed. Therefore, the SWICS 1.1 and SWICS 2.0 datasets should not be directly combined for long-term trend analysis. In this study, we strictly separate the data before and after the instrument anomaly to avoid the risk of data mixing. We analyzed the two time periods separately. The anomaly occurred on August 23, 2011, so we define 1999-2009 as the pre-anomaly period and 2012-2020 as the post-anomaly period, which broadly cover the main phases of SC 23 and SC 24 while effectively avoiding data mixing.

\section{Results of all two sample K-S tests.}
\label{sec:appB}
Table \ref{tab4} summarizes all the results of the two sample K-S tests. The first column lists the parameters. The second column indicates the data subset used, either by time period or by solar wind type. The third column specifies the two groups being compared in each test. The fourth column shows the H values, the fifth column shows the P values, the sixth column shows the D values, and the seventh column shows the $D_{\mathrm{crit}}$ values. The significance level used is 0.05.
\begin{table}[!htbp]
\tiny
\renewcommand{\arraystretch}{0.65}
\centering
\caption{Results of all two sample K-S tests.}
\label{tab4}
\begin{tabular}{ccccccc}
\midrule
\midrule
  Parameter                                        &  Data Subset              &  Groups Compared            & H & P         & D      & Dcrit  \\
\midrule
\multirow{3}{*}{latitude}                          & $\alpha$                         & SC 23 vs SC 24 & 0 & 0.94065   & 0.0365 & 0.0941 \\
\cmidrule(r){2-7}
                                                   & $\beta$                         & SC 23 vs SC 24 & 0 & 0.24265   & 0.0388 & 0.0516 \\
\cmidrule(r){2-7}
                                                   & complex                   & SC 23 vs SC 24 & 0 & 0.083668  & 0.1038 & 0.1131 \\
\midrule
\multirow{6}{*}{footpoint magnetic field strength} & \multirow{3}{*}{ALL TIME} & $\alpha$ vs $\beta$         & 1 & 2.39E-144 & 0.2397 & 0.0247 \\
                                                   &                           & $\alpha$ vs complex   & 1 & 0  & 0.389  & 0.0256 \\
                                                   &                           & $\beta$ vs complex   & 1 & 5.59E-106 & 0.1633 & 0.0193 \\
\cmidrule(r){2-7}
                                                   & $\alpha$                         & SC 23 vs SC 24 & 1 & 5.01E-27  & 0.1764 & 0.0424 \\
\cmidrule(r){2-7}
                                                   & $\beta$                         & SC 23 vs SC 24 & 1 & 5.75E-54  & 0.1613 & 0.0272 \\
\cmidrule(r){2-7}
                                                   & complex                   & SC 23 vs SC 24 & 1 & 2.23E-49  & 0.1765 & 0.0304 \\
\midrule
\multirow{6}{*}{in-situ magnetic field strength}   & \multirow{3}{*}{ALL TIME} & $\alpha$ vs $\beta$         & 1 & 1.90E-28  & 0.1040 & 0.0247 \\
                                                   &                           & $\alpha$ vs complex   & 1 & 5.74E-47  & 0.1390 & 0.0256 \\
                                                   &                           & $\beta$ vs complex   & 1 & 1.36E-11  & 0.0515 & 0.0193 \\
\cmidrule(r){2-7}
                                                   & $\alpha$                         & SC 23 vs SC 24 & 1 & 2.60E-08  & 0.0944 & 0.0424 \\
\cmidrule(r){2-7}
                                                   & $\beta$                         & SC 23 vs SC 24 & 1 & 9.02E-24  & 0.1046 & 0.0272 \\
\cmidrule(r){2-7}
                                                   & complex                   & SC 23 vs SC 24 & 1 & 7.20E-55  & 0.1786 & 0.0304 \\
\midrule
\multirow{6}{*}{speed}                             & \multirow{3}{*}{ALL TIME} & $\alpha$ vs $\beta$         & 1 & 8.50E-14  & 0.0719 & 0.0247 \\
                                                   &                           & $\alpha$ vs complex   & 1 & 2.12E-05  & 0.0455 & 0.0256 \\
                                                   &                           & $\beta$ vs complex   & 1 & 1.04E-19  & 0.0678 & 0.0193 \\
\cmidrule(r){2-7}
                                                   & $\alpha$                         & SC 23 vs SC 24 & 1 & 1.64E-53  & 0.2453 & 0.0424 \\
\cmidrule(r){2-7}
                                                   & $\beta$                         & SC 23 vs SC 24 & 1 & 9.69E-81  & 0.1942 & 0.0272 \\
\cmidrule(r){2-7}
                                                   & complex                   & SC 23 vs SC 24 & 1 & 2.99E-21  & 0.1103 & 0.0304 \\
\midrule
\multirow{9}{*}{$O^{7+}/O^{6+}$}                           & \multirow{3}{*}{SC 23}    & $\alpha$ vs $\beta$         & 1 & 1.79E-22  & 0.1314 & 0.0350 \\
                                                   &                           & $\alpha$ vs complex   & 1 & 1.83E-98  & 0.2857 & 0.0361 \\
                                                   &                           & $\beta$ vs complex   & 1 & 2.31E-83  & 0.1750 & 0.0238 \\
\cmidrule(r){2-7}
                                                   & \multirow{3}{*}{SC 24}    & $\alpha$ vs $\beta$         & 1 & 1.59E-13  & 0.1053 & 0.0363 \\
                                                   &                           & $\alpha$ vs complex   & 1 & 5.08E-23  & 0.1455 & 0.0377 \\
                                                   &                           & $\beta$ vs complex   & 1 & 3.49E-06  & 0.0656 & 0.0331 \\
\cmidrule(r){2-7}
                                                   & $\alpha$                         & SC 23 vs SC 24 & 1 & 3.99E-42  & 0.2181 & 0.0424 \\
\cmidrule(r){2-7}
                                                   & $\beta$                         & SC 23 vs SC 24 & 1 & 6.04E-43  & 0.1441 & 0.0272 \\
\cmidrule(r){2-7}
                                                   & complex                   & SC 23 vs SC 24 & 1 & 1.21E-79  & 0.2230 & 0.0304 \\
\midrule
\multirow{9}{*}{$Q_{Fe}$}                               & \multirow{3}{*}{SC 23}                     & $\alpha$ vs $\beta$         & 1 & 5.07E-06  & 0.0663 & 0.0350 \\
                                                   &                           & $\alpha$ vs complex   & 1 & 9.15E-11  & 0.0929 & 0.0361 \\
                                                   &                           & $\beta$ vs complex   & 1 & 5.53E-16  & 0.0758 & 0.0238 \\
\cmidrule(r){2-7}
                                                   & \multirow{3}{*}{SC 24}                     & $\alpha$ vs $\beta$         & 1 & 0.019172  & 0.0413 & 0.0363 \\
                                                   &                           & $\alpha$ vs complex   & 1 & 6.67E-12  & 0.1037 & 0.0377 \\
                                                   &                           & $\beta$ vs complex   & 1 & 2.20E-11  & 0.0905 & 0.0331 \\
\cmidrule(r){2-7}
                                                   & $\alpha$                         & SC 23 vs SC 24 & 1 & 3.97E-38  & 0.2076 & 0.0424 \\
\cmidrule(r){2-7}
                                                   & $\beta$                         & SC 23 vs SC 24 & 1 & 2.78E-61  & 0.1725 & 0.0272 \\
\cmidrule(r){2-7}
                                                   & complex                   & SC 23 vs SC 24 & 1 & 1.72E-46  & 0.1701 & 0.0304 \\
\midrule
\multirow{6}{*}{$A_{He}$}                               & \multirow{3}{*}{ALL TIME} & $\alpha$ vs $\beta$         & 1 & 2.11E-03  & 0.0346 & 0.0247 \\
                                                   &                           & $\alpha$ vs complex   & 1 & 6.99E-24  & 0.1006 & 0.0256 \\
                                                   &                           & $\beta$ vs complex   & 1 & 7.01E-25  & 0.0779 & 0.0193 \\
\cmidrule(r){2-7}
                                                   & $\alpha$                         & SC 23 vs SC 24 & 1 & 0.0066126 & 0.0542 & 0.0424 \\
\cmidrule(r){2-7}
                                                   & $\beta$                         & SC 23 vs SC 24 & 1 & 2.59E-34  & 0.1288 & 0.0272 \\
\cmidrule(r){2-7}
                                                   & complex                   & SC 23 vs SC 24 & 1 & 2.42E-14  & 0.0919 & 0.0304 \\
\midrule
\multirow{9}{*}{FIP bias}                          & \multirow{3}{*}{SC 23}    & $\alpha$ vs $\beta$         & 1 & 2.07E-26  & 0.1435 & 0.0350 \\
                                                   &                           & $\alpha$ vs complex   & 1 & 1.56E-45  & 0.1948 & 0.0361 \\
                                                   &                           & $\beta$ vs complex   & 1 & 7.09E-14  & 0.0706 & 0.0238 \\
\cmidrule(r){2-7}
                                                   & \multirow{3}{*}{SC 24}    & $\alpha$ vs $\beta$         & 1 & 0.0042805 & 0.0475 & 0.0363 \\
                                                   &                           & $\alpha$ vs complex   & 1 & 2.41E-39  & 0.1909 & 0.0377 \\
                                                   &                           & $\beta$ vs complex   & 1 & 1.05E-39  & 0.1713 & 0.0331 \\
\cmidrule(r){2-7}
                                                   & $\alpha$                         & SC 23 vs SC 24 & 1 & 0         & 0.6453 & 0.0424 \\
\cmidrule(r){2-7}
                                                   & $\beta$                         & SC 23 vs SC 24 & 1 & 0         & 0.4833 & 0.0272 \\
\cmidrule(r){2-7}
                                                   & complex                   & SC 23 vs SC 24 & 1 & 0         & 0.5310 & 0.0304 \\
\midrule
\end{tabular}
\end{table}

\end{appendix}

\end{document}